\documentclass[twocolumn]{aastex6}  

\usepackage{graphicx}
\usepackage{placeins} 
\usepackage{listings}
\usepackage[title]{appendix}

\usepackage{amsmath}
\usepackage{xcolor}

\shorttitle{GENGA II}
\shortauthors{Grimm et Al}

\begin{document}

\title{GENGA. II. GPU planetary $N$-body simulations with non-Newtonian forces and high number of particles.}

\author{Simon L. Grimm\altaffilmark{1}}
\author{Joachim G. Stadel\altaffilmark{2}}
\author{Ramon Brasser\altaffilmark{3}}
\author{Matthias M. M. Meier\altaffilmark{4,5}}
\author{Christoph Mordasini\altaffilmark{6}}

\altaffiltext{1}{University of Bern, Center for Space and Habitability, Gesellschaftsstrasse 6, CH-3012, Bern, Switzerland.  Email: simon.l.grimm@unibe.ch}

\altaffiltext{2}{University of Z\"{u}rich, Institute for Computational Science, Winterthurerstrasse 190, CH-8057, Z\"{u}rich, Switzerland. }

\altaffiltext{3}{Origins Research Institute, Centre for Astronomy and Earth Sciences, Konkoly Thege Miklos St 15-17, H-1121 Budapest, Hungary.}

\altaffiltext{4}{Naturmuseum St.Gallen, Rorschacher Strasse 263, 9016 St.Gallen, Switzerland. Email: matthias.meier@naturmuseumsg.ch}
\altaffiltext{5}{ETH Zurich, Institute of Geochemistry and Petrology, Clausiusstrasse 25, 8092 Zurich, Switzerland.}

\altaffiltext{6}{University of Bern, Physikalisches Institut, Gesellschaftsstrasse 6, CH-3012, Bern, Switzerland.}


\begin{abstract}
We present recent updates and improvements of the graphical processing unit (GPU) $N$-body code GENGA. Modern state-of-the-art simulations of planet formation require the use of a very high number of particles to accurately resolve planetary growth and to quantify the effect of dynamical friction. At present the practical upper limit is in the range of 30,000 - 60,000 fully interactive particles; possibly a little more on the latest GPU devices. While the original hybrid symplectic integration method has difficulties to scale up to these numbers, we have improved the integration method by i) introducing higher level changeover functions and ii) code improvements to better use the most recent GPU hardware efficiently for such large simulations. We added treatments of non-Newtonian forces such as general relativity, tidal interaction, rotational deformation, the Yarkovsky effect, and Poynting-Robertson drag, as well as a new model to treat virtual collisions of small bodies in the solar system. We added new tools to GENGA, such as semi-active test particles that feel more massive bodies but not each other, a more accurate collision handling and a real-time openGL visualization. We present example simulations, including a 1.5 billion year terrestrial planet formation simulation that initially started with 65,536 particles, a 3.5 billion year simulation without gas giants starting with 32,768 particles, the evolution of asteroid fragments in the solar system, and the planetesimal accretion of a growing Jupiter simulation. GENGA runs on modern NVIDIA and AMD GPUs.
\end{abstract}


\section{Introduction}
GENGA is an $N$-body code optimized for simulating planet formation and planetary system evolution. It runs on graphical processing units (GPUs) to speed up the integration time. On average, a high-end GPU has a similar amount of computing power as a 128-core CPU, but with less than half of the power consumption \citep{PortegiesZwart2020}. Since the release of the first version of GENGA \citep{GrimmStadel2014}, available computational resources have improved dramatically in performance, but especially GPU performance. On the other hand, the requirements of state-of-the-art simulations in planet formation have also increased dramatically. Current hardware allows for 30,000 - 60,000 fully interactive planetesimals to be simulated for 10 Myr in about 60 days. Examples of these high-end simulations are given in \cite{Quarles2019}, \cite{Clement+2020} and \cite{Woo2021}. Since the parallel programming environment in the CUDA language has evolved, it is also necessary to reconsider some of the parallelization techniques used in GENGA and to optimize or adapt them further to suit modern hardware.

In addition to a good performance, a modern $N$-body code for planet formation or long-term planetary system evolution must also include non-Newtonian forces. General relativity corrections, tidal forces and rotational deformation forces are necessary to simulate the long-term behavior of compact exoplanetary systems. In simulations including small bodies, the Yarkovsky effect and eventually Poynting-Robertson drag must be implemented to calculate volatile material between different parts of the planetary system more realistically.

Examples of $N$-body CPU codes including such additional forces are Mercury-T \citep{Bolmont2015} or Posidonius \citep{Blanco-Cuaresma+Bolmont}.
Codes that used CPU parallel $N$-body methods are REBOUND \citep{ReinLiu2012} and PKDGRAV3 \citep{PotterStadelTeyssier2017}, while GPUs are used in the codes QYMSYM \citep{MooreQuillen2010}, GLISSE \citep{Zhang+2022}, or Hiperion\footnote{https://www.konkoly.hu/staff/regaly/research/hiperion.html}.

GENGA uses a hybrid symplectic integration method, which is able to maintain a good long-term energy conservation while it is also able to resolve close-encounters accurately. This is possible by using a smooth changeover function that moves the gravitational forces between close-encounter pairs from the symplectic integrator to a direct $N$-body solver. By increasing the number of particles in the simulations, it has become evident that this integration method does not scale up well. Since the number of close-encounters can grow very fast with the number of particles used, the efficiency of the parallelization can be affected. This fact makes evident the need to reconsider the hybrid symplectic integration method for large $N$-simulations. In this work, we introduce higher level changeover functions and we demonstrate how this new method is able to perform simulations with up to $\sim$60,000 fully interactive planetesimals.

In addition to non-Newtonian forces and an improved integration method, we add a variety of new options and tools to GENGA, which we describe below. This paper is structured as follows:

In section \ref{sect:HybridSymplectic}, we summarize the theory of the hybrid symplectic integrator and explain why the integration method needs to be improved in order to handle a large number of close-encounters. We continue with a detailed description of newly added non-Newtonian forces. These are general relativity corrections in section \ref{sect:GR}, tidal forces and rotational deformation forces in section \ref{sect:Tides+Rot}. Both of these sections contain a testing and comparison subsection. In section \ref{sect:smallBodies}, we introduce new forces and models for small-body simulations. These are the Yarkovsky effect in section \ref{sect:Yarkovsky}, Poynting-Robertson drag in section \ref{sect:PR}, a collisional break-up model for meteoroid dynamics of the solar system in section \ref{Collision} and a model for collisional induced rotation-rate reset in section \ref{CollisionalReset}. The latter two models are useful for example when the transport of small ejecta material in the solar system is studied. We present an example simulation of ejecta material from the asteroid 6 Hebe in section \ref{sect:Hebe}.

After introducing new forces and models we continue with a detailed description of improvements of the integration method in section \ref{sect:closeEncounters}. These include more efficient implementations of the Bulirsch-Stoer integration kernels needed for the close-encounter handling presented in section \ref{sect:BS}, and the introduction of a higher level changeover function shown in section \ref{sect:SLevels}.
In the sections \ref{sect:ExampleI} and \ref{sect:ExampleII} we present two long-term example simulations of terrestrial planet formation with and without gas giants and a gas disk. For both simulations, the results are shown for different numbers of initial particles, ranging from 2048 to 65,536.

In section \ref{sect:Coll+Enc}, we describe improvements in the collision and encounter handling. These include the option of a more precise collision resolution, the option for reporting and tracing back collisions at a time before they really collide, and the option to report all close-encounter events during a simulation. New tools in GENGA are described in section \ref{sect:NewTools}. These include a new semi-active test particle mode in section \ref{sect:TP2} and the option of using predefined coordinates, mass, or radii tracks in section \ref{sect:SetElements}.

Finally, the performance of GENGA is analyzed in section \ref{sect:Performance}.

GENGA is open source and available at https://bitbucket.org/sigrimm/genga .
The repository also contains a detailed user documentation and a tutorial where GENGA can be tested online.

\section{Theory of the hybrid symplecic integrator}
\label{sect:HybridSymplectic}

A symplectic integration method \citep[e.g.][]{WisdomHolman1991} is able to integrate planetary orbits over billions of years without a drift in energy and angular momentum. Since the method is restricted to use a fixed time step, it is difficult to handle close-encounters between bodies where the time step of the involved bodies must be reduced substantially. It also presents a numerical challenge if the orbital periods in the system vary over several orders of magnitude, since the orbit of the innermost bodies determine the overall time step. The hybrid symplectic method developed by \citet{Chambers99} uses a smooth changeover function to transfer the calculation of close-encounters from the symplectic to a direct $N$-body integrator. We also use a Bulirsch-Stoer \citep{BulirschStoer2002, Press2007} method to resolve the close-encounter part, but other adaptive time step $N$-body methods could be used here as well, such as a Hermite integrator \citep{Makino+1992} or embedded Runge-Kutta schemes \citep{Hairer1993} even if these may cause artificial precession in the orbits. The transition between the symplectic part and the direct $N$-body part must be applied smoothly enough to prevent the accumulation of large errors in the total energy. Therefore, a critical radius must be defined to set a threshold between the close-encounter regime and the normal integration regime. It must be chosen to be large enough to ensure a smooth transition, but small enough to avoid spending too much simulation time in the encounter phase. In the following section, we repeat the most important parts of the hybrid symplectic integrator. For more details, we refer to \cite{Chambers99} or \cite{GrimmStadel2014}.

By using democratic coordinates (heliocentric positions and barycentric velocities), the Hamiltonian system of a planetary system can be split into three parts $H_A$, $H_B$ and $H_C$, which correspond to the Keplerian part, the interaction part, and the Sun part of the Hamiltonian system, respectively. These are given by
\begin{eqnarray}
\label{H}
 H = H_{A} + H_{B} + H_{C},
\end{eqnarray}

with 
\begin{eqnarray}
\label{Ha}
H_{A} = \sum_{i=1}^{N} \left( \frac{p_{i}^{2}}{2m_{i}}  - \frac{G m_{i} m_{\star}}{r_{i\star}} \right) \nonumber \\
- \sum_{i = 1}^{N} \sum_{j = i+1}^{N} \frac{G m_{i} m_{j}}{r_{ij}} [ 1 - K(r_{ij})],
\end{eqnarray}

\begin{equation}
\label{Hb}
H_{B} = -\sum_{i = 1}^{N}\sum_{j=i +1}^{N} \frac{G m_{i} m_{j} }{r_{ij}} K(r_{ij})
\end{equation}
and
\begin{equation}
\label{Hc}
 H_{C} = \frac{1}{2m_{\star}}\left( \sum _{i =1} ^{N} \mathbf{p}_{i} \right) ^{2},
\end{equation}

where the symbol $\star$ refers to the central mass and $K(r_{ij})$ is a smooth changeover function ranging from 0 to 1 (see top panel of Figure \ref{fig:K}). The distance between the particles $i$ and $j$ is $r_{ij}$.

The limit where the changeover functions is applied is defined by a critical radius $r_{\text{crit}}$ of a particle $i$, which is computed as

\begin{equation}
    \label{rcrit}
    r_{\text{crit},i}= \max(n1 \cdot R_{H,i}, n2 \cdot dt \cdot v_i).
\end{equation}

The critical radius depends on two terms. The first depends on the Hill radius $R_H$ while the second contains the time step $dt$ and the Kepler velocity $v$ of the particle $i$. The two parameters $n1$ and $n2$ are constants. Experimentation has found that typical values that yield a good compromise between accuracy and computation time are $n_1=3$ and $n_2=0.4$ \citep{Chambers99}.

The integrator needs to search for close-encounter pairs at each time step and to sort them into independent close-encounter groups. These groups are then integrated with the Bulirsch-Stoer direct $N$-body method \citep{GrimmStadel2014}. Ideally, the close-encounter groups consist of only a single pair of bodies, but it can happen that bodies have multiple close-encounter pairs, which need to be linked together in a bigger close-encounter group. In the worst case scenario, all bodies are in a close-encounter with some neighboring bodies, and all of them are linked together into a single giant close-encounter group. This scenario is likely to happen when the particle number density is increased during high-$N$ simulations. In that case, the original hybrid symplectic method becomes very inefficient and must be improved or replaced with something else. We describe this in section \ref{sect:SLevels}.

\subsection{High-$N$ simulations and large close-encounter groups}
The critical radius $r_{\text{crit},i}$ depends directly on the velocity of the particles, which introduces a difficulty when running high-resolution simulations with a large number of small planetesimals. While the first term in the critical radius decreases as the Hill radius decreases, the second term does not depend on the particle radii or masses. This means that, by increasing the particle surface or volume density of the simulations, the critical radii of the planetesimals do not change much and more and more particles are considered to be in a mutual close-encounter regime. Ultimately this can lead to a situation whereby every particle is in a close-encounter with its neighbor, and therefore all particles are linked together into one very large close-encounter group, as illustrated in Figure \ref{fig:ring}. This behavior is already described in \citet[section 5]{GrimmStadel2014}.

The original version of GENGA only permitted close-encounter group sizes of at most 128 particles because of memory limitations in the first Bulirsch-Stoer implementation of \cite{GrimmStadel2014}. In section \ref{sect:closeEncounters} we present a new implementation without an upper limit on the close-encounter group sizes. However, the performance of these very large close-encounter groups is less efficient than small groups and this situation should still be avoided if possible. In section \ref{sect:SLevels}, we present a way to reduce the size of close-encounter groups even for high-resolution simulations.

\subsection{Non-Newtonian forces}
Celestial mechanics contains a variety of different forces. Besides the dominant Newtonian gravitational force, thermal forces, tides, or general relativity effects can be important. While \cite{GrimmStadel2014} already contains a model for a gas disk, we add in this work the most important non-Newtonian forces to GENGA.  Other forces can be added individually. The user can enable or disable all these forces in the GENGA parameter file.

In general, these additional forces can depend on both the positions and the velocities of the bodies, which would violate the symplectic structure of the integrator because these forces would not be separable in the Hamiltonian system \citep{SahaTremaine1994}. Applying such a force in a usual kick operation would lead to an error in the energy which should be avoided \citep{SahaTremaine1994}. A simple solution to this problem is to calculate the force with an implicit midpoint method because it is also symplectic \citep{SahaStadelTremaine1997}.

In sections \ref{sect:GR}-\ref{sect:smallBodies}, we describe the newly added forces in detail. These are general relativity corrections, tidal forces, rotational deformation, the Yarkovsky effect, and Poynting-Robertson drag. These forces are expressed in heliocentric coordinates. That means that before the forces can be applied in the integration, barycentric velocities must be converted to heliocentric coordinates and back after the non-Newtonian force step, which causes computational overhead.

\subsection{The implicit midpoint method}
\label{sect:implicit}
A conservative implementation of a general force $f$, which is not separable in the Hamiltonian system, can be achieved by the implicit midpoint method, which is given by \citep{Hairer1993}
\begin{equation}
\label{eqn:implicit}
\mathbf{v}^{t + dt}_i = \mathbf{v}^{t}_i + dt \cdot \mathbf{a}_{f}\left(\mathbf{r}^t_i, \frac{\mathbf{v}^{t}_i + \mathbf{v}^{t + dt}_i}{2} \right).
\end{equation}
Here $\mathbf{a}_{f}$ is the acceleration caused by the force $f$. Equation \ref{eqn:implicit} needs to be iterated until the difference between $\mathbf{v}^{t + dt}_i$ and $\mathbf{v}^{t}_i$ is smaller than a defined tolerance value. In practise, we find that fewer than three iterations are needed.
We use the implicit midpoint method in GENGA to implement non-Newtonian forces that are not separable in the Hamiltonian.

\section{General Relativity}
\label{sect:GR}
General relativity (GR) causes a perihelion shift of all bodies, orbiting a central mass as well as an increase in the mean motion \citep{SahaTremaine1994}. Due to energy conservation the semi major axis is not affected and remains constant. While the effect of GR is usually small for solar system formation simulations, it can play an important role for objects close to the central star such as compact systems of exoplanets. This is especially true for long-term stability analysis of planetary systems.

GR can be approximated by post-Newtonian corrections of the form \citep{Kidder1995, MardlingLin2002, Fabrycky2010}:

\begin{eqnarray}
\label{eqn:apn}
\begin{aligned}
\mathbf{a}_{PN} = -\frac{G(M_{\star} + m_i)}{r^2_i c^2} \cdot 
\bigg\{ -2(2 - \eta) \dot{r}_i \mathbf{v}_i \\
+ \left[ (1 + 3 \eta) \mathbf{v}_i \cdot \mathbf{v}_i 
- \frac{3}{2} \eta \dot{r}^2_i - 2(2+\eta) \frac{G(M_{\star} + m_i)}{r_i}\right] \mathbf{\hat{r}}\bigg\} , 
 \end{aligned}
\end{eqnarray}

where $c$ is the speed of light and $\eta$ is defined as:
\begin{eqnarray}
 \eta = \frac{M_{\star}m_i}{(M_{\star} + m_i)^2}.
\end{eqnarray}
A similar description is given in \cite{Sitarski1983} or at a higher order in \citet[Equation 4-26]{Moyer2003}. Since equation \ref{eqn:apn} depends on the positions and the velocities, the GR corrections must by applied with the implicit midpoint method described above in section \ref{sect:implicit}.

\subsection{GR Hamiltonian splitting}
As an alternative to the previous formulation, \cite{SahaTremaine1994} provide a description of general relativity corrections by using a Hamiltonian splitting approach. The GR (or also called post-Newtonian) Hamiltonian equation takes the form given by

\begin{equation}
 H_{PN} = \sum_i \left( \alpha_i H_{Kep,i}^2 + \frac{\beta_i}{r^2_i} + \gamma_i p^4_i \right),
 \end{equation}
 with the quantities
\begin{eqnarray} 
 \alpha_i = \frac{3}{2m_i c^2}, \\ \nonumber
 \beta_i = \frac{-\mu_i^2 m_i}{c^2},  \\ \nonumber
 \gamma_i = -\frac{1}{2m_i^3 c^2}, \\ \nonumber
 \mu_i = G(M_\star + m_i),
\end{eqnarray}
and the speed of light $c$.

Following \cite{SahaTremaine1994} the term $\gamma_i p^4_i$ in the Hamiltonian equation can be expressed as
\begin{equation}
d \mathbf{x}_i = -dt \cdot 2 \frac{v^2_i}{c^2} \mathbf{v}_i,
\end{equation}
with the time step $dt$. This term can be combined with the $H_C$ term in equation \ref{Hc}.
The $\beta$ term in the Hamiltonian equation can be expressed as
\begin{equation}
d \mathbf{v}_i = -dt \cdot 2 \frac{\mu^ 2}{c^2 r^4_i} \mathbf{r}_i,
\end{equation}
and can be combined with the $H_B$ term in equation \ref{Hb}. However, care must be taken that the changeover function $K(r_{ij})$ does not affect this term.

Following \citet{SahaTremaine1994} yet again, the $\alpha$ term can be implemented through a time step modification in the particle drift operation (equation \ref{Ha}):
\begin{equation}
\label{eqn:pseudo}
dt_i' = dt \cdot \left( 1 - \frac{3}{2} \frac{\mu}{a_i c^ 2}\right).
\end{equation}
This time step modification must be applied in both the FG function of the Kepler solver \citep{GrimmStadel2014} as well as the Bulirsch-Stoer direct solver for close-encounters. It is important to only modify the acceleration terms between the central mass and the particles, $\mathbf{a}_{i \star}$, and {\it not} between two particles, $\mathbf{a}_{ij}$.

An important detail is that the formulation above requires the use of pseudovelocities. \cite{SahaTremaine1994} recommend to carry out the entire integration in pseudovelocities and only convert back to real velocities at integration output times. However, GENGA allows the use of other non-Newtonian forces that can also depend on the velocities. We cannot make use of this approach and instead we are required to compute all the GR terms and other forces in real velocities rather than in pseudovelocities. Therefore, at each time step, we convert from real velocities to pseudovelocities and back, causing a slight computational overhead. Only the Sun-kick and the drift operation is done in pseudovelocities. Since the conversion in Equation \ref{eqn:pseudo} can be done for each particle individually, the parallelization of it is trivial. However, this approach requires two additional CUDA kernel calls, which reduces the performance of small simulations because these are not dominated by the $N^2$ kick operations or by close-encounters.

\subsection{Comparison and test of the GR effect}
\begin{figure}
\includegraphics[width=1.0\linewidth]{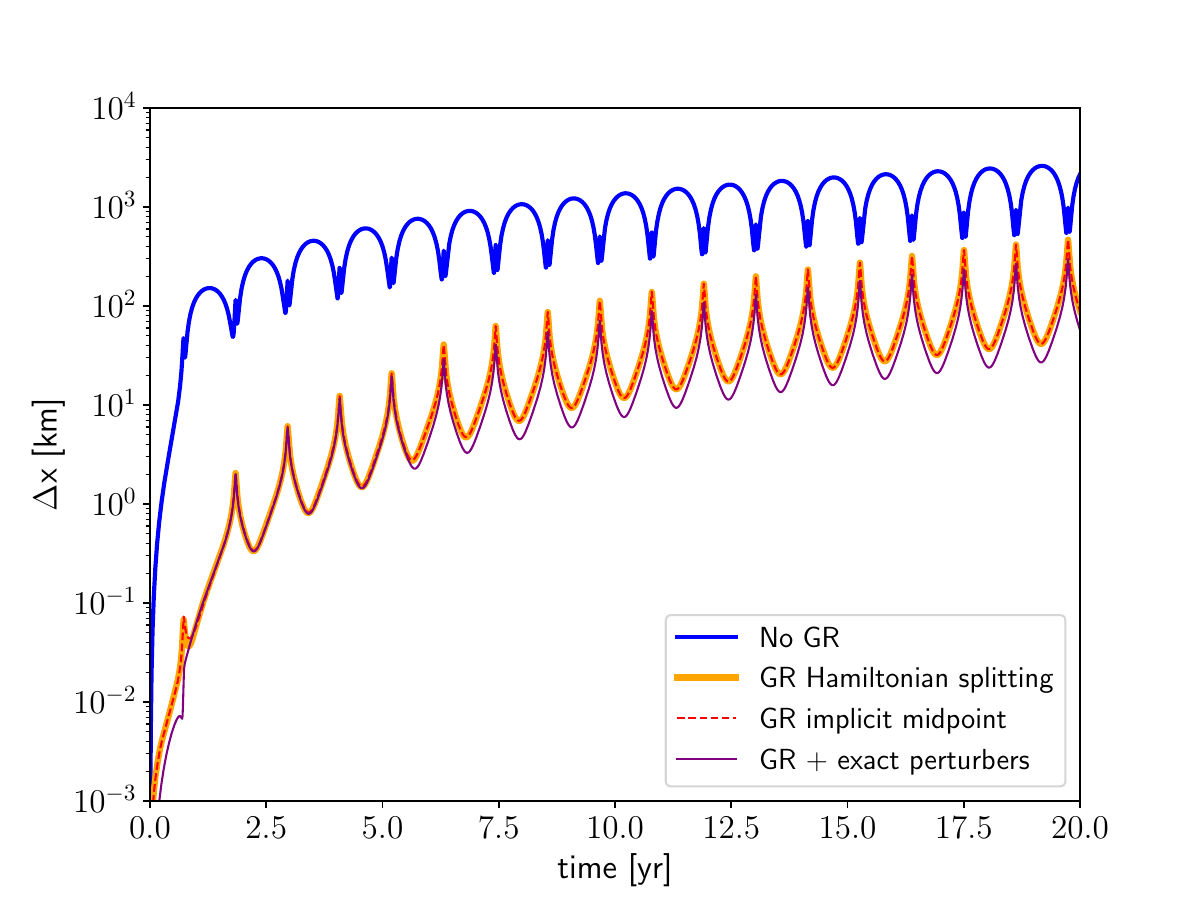}
\caption{Difference in position of the asteroid (1566) Icarus between the integration and the measured orbit. The Asteroid Icarus is integrated together with the 25 most massive objects of the solar system. When GR corrections are not considered (blue line) then the position after two orbital periods differs by more than 100 km. With GR effects enabled, the difference is less than 2 km. There is not significant difference between the Hamiltonian splitting approach (orange line)  and the implicit midpoint method (red dashed line). When the position of the 25 perturbers is not integrated, but interpolated from their measured positions, then the difference in position is reduced further (purple line).
}
\label{fig:GRdiff} 
\end{figure}

A good object to test the general relativity effect is evolution of the asteroid (1566) Icarus \citep{Shapiro1971, Sitarski1992}. It has an orbital eccentricity of 0.83 so that it ventures closer than 0.2 au to the Sun. In order to compute the evolution of the orbit of Icarus, we integrate it together with the 25 most massive objects of the solar system. All data are taken from the JPL HORIZON system \footnote{https://ssd.jpl.nasa.gov/horizons.cgi}. We compare our integration with the measured orbit of Icarus. The results are shown in Figure \ref{fig:GRdiff}. One can clearly see how the GR corrections improve the orbital positions during the first few orbital periods. Comparing the Hamiltonian splitting approach with the implicit midpoint method leads to no significant difference. The integration precision can be improved when the 25 perturber objects are not integrated, but rather interpolated from the known ephemeris. The reason for this is that in reality other forces and effects occur, which we do not include here and which can influence the positions of the perturbers slightly. 
The remaining difference in the orbital position after applying the GR corrections is most likely caused by other non-gravitational effects \citep{Greenberg2017}, and small deviations in the initial conditions increase after each revolution. 

The GR effect calculation scales linearly with the number of particles $N$, so that the performance of large simulations ($N \gtrapprox 4096$) is only affected marginally. Simulations with $N \approx 1024$ are slowed down by $\sim$ $1\%$, while small simulations ($N \lesssim 128$) are affected more. A simulation with 16 fully interacting bodies can take up to 1.5 times longer with the implicit midpoint method, and more than twice as long with the Hamiltonian splitting method. The user can choose between the two implementations with a setting in the GENGA parameter file. We recommend using the implicit midpoint method, since it gives a better performance for similar accuracy. 

\section{Tidal forces and rotational deformation}
\label{sect:Tides+Rot}

In this section, we describe the implementation of the tidal force and the rotational deformation force. Both of these forces depend on the positions and the velocities, therefore they are calculated with the implicit midpoint method, described in section \ref{sect:implicit}.
All equations are given in heliocentric coordinates.
\subsection{Tidal forces}
\label{sect:tides}
For implementing the tidal effects, we follow the weak friction tidal model described in \cite{Hut1981} and \cite{Bolmont2015}. In this model, the tidal bulge of a primary body with radius $R$, induced by a companion body of mass $m$, is described by two point masses at the surface of the primary body with mass \citep{Hut1981} 
\begin{equation}
\mu \approx \frac{1}{2} k_{2,i} m R^3 \left[r(t - \tau) \right] ^{-3},
\end{equation}
where $r(t - \tau)$ is the distance between the two bodies at time $t - \tau$, and $\tau$ is a constant time lag, which models the tidal dissipation. The quantity $k_{2,i}$ is the potential Love number of degree 2 of the body $i$. Introducing the  quantities
\begin{equation}
D_{to,i} = 3G\frac{m^2_{\star}R^5_i}{r^7_i} k_{2,i}
\end{equation}
\begin{equation}
D_{to,\star} = 3G\frac{m^2_i R^5_\star}{r^7_i} k_{2,\star}
\end{equation}

leads to equation 5 in \cite{Bolmont2015}, which is given by
\begin{equation}
P_{to,i} = D_{to,i} \tau_i
\end{equation}
\begin{equation}
P_{to,\star} = D_{to,\star} \tau_{\star}.
\end{equation}

The radial part of the tidal force can be written as
\begin{eqnarray}
F_{Tr} = - (D_{to,\star} + D_{to,i}) -3\frac{\dot{r}_i}{r_i}  (P_{to,\star} + P_{to,i}),
\end{eqnarray}
where $r_i$ is the distance between body $i$ and the central mass $\star$, and $\dot{r}_i = \frac{dr_i}{dt}$.

The total tidal force can be written as equation 6 in \cite{Bolmont2015}, where we corrected a typo in the first line (also corrected in \cite{Bolmont+2020}), which is given by

\begin{eqnarray}
\mathbf{F}_T = F_{Tr}\mathbf{e}_{ri} \\ \nonumber
+ P_{to,i} \left(\mathbf{\Omega}_i \times \mathbf{e}_{ri} \right) +  P_{t0,\star} \left(\mathbf{\Omega}_{\star} \times \mathbf{e}_{ri} \right) \\ \nonumber
- \left(P_{to,i} +  P_{to,\star} \right)\left(\mathbf{\dot{\theta}}_i \times \mathbf{e}_{ri} \right),
\end{eqnarray}

where $\mathbf{e}_{ri}$ is the radial unit vector, and $\mathbf{\Omega}_{\star}$ is the rotational angular velocity vector of the star, $\mathbf{\Omega}_i$ the rotational angular velocity vector of body $i$, and $\mathbf{\dot{\theta}}_i$ describes the orbital velocity of the body $i$. A figure explaining the tidal force model is given in \cite{Hut1981} and in \cite{Bolmont2015}.

Rewriting the last term as
\begin{eqnarray}
\left(\mathbf{\dot{\theta}}_i \times \mathbf{e}_{ri} \right) = \frac{1}{r^3_i} \left(  \mathbf{r}_i \times \mathbf{v}_i\right)  \times \mathbf{r}_i \\ \nonumber
= \frac{\mathbf{v}_i}{r_i} - \frac{(\mathbf{r}_i \cdot \mathbf{v}_i)\mathbf{r}_i }{r^3_i} \\ \nonumber
= \frac{\mathbf{v}_i}{r_i} - \frac{\dot{r}_i}{r_i} \mathbf{e}_{ri},
\end{eqnarray}
leads to
\begin{eqnarray}
\mathbf{F}_T = \left[- (D_{to,\star} + D_{to,i}) -2\frac{\dot{r}_i}{r_i}  (P_{to,\star} + P_{to,i})\right] \mathbf{e}_{ri} \\ \nonumber
+ P_{to,i} \left(\mathbf{\Omega}_i \times \mathbf{e}_{ri} \right) +  P_{to,\star} \left(\mathbf{\Omega}_{\star} \times \mathbf{e}_{ri} \right) \\ \nonumber
- \left(P_{to,i} +  P_{to,\star} \right) \frac{\mathbf{v}_i}{r_i}.
\end{eqnarray}

Note that there are different models of the tidal forces in existence other than the constant time lag description that we use. For more details see, e.g., \cite{Efroimsky2007}, \cite{Leconte+2010}, or \cite{Auclair-Desrotour+2014}. Some of these models can also be treated in a secular evolution model. In section \ref{sect:tidalTest} we compare our implementation to three secular evolution models.
\cite{Leconte+2010} describe how the time lag $\tau$ can be converted into the tidal quality factor $Q$. While there is no simple relation for the general case, the case of a non-synchronized circular orbit with $Q \gg 1$ can be converted as \citep{Leconte+2010}
\begin{equation}
    Q = \frac{1}{2 \tau |\omega - n|}.
\end{equation}

\subsection{Tidal torque}
In addition to the acceleration on the particles the tidal force generates a torque, which changes the spin angular momentum of the particles and of the central star. This tidal torque is given as \citep{Hut1981}
\begin{equation}
    \mathbf{N}_T = \mathbf{r} \times \mathbf{F}_{T\theta},
\end{equation}
where $\mathbf{F}_{T\theta}$ is the transverse component of the tidal force. With the relation
\begin{equation}
    \mathbf{r} \times (\mathbf{\Omega} \times \mathbf{e}_r) - \mathbf{r} \times \frac{\mathbf{v}}{r} = 
    \mathbf{\Omega}r - (\mathbf{r} \cdot \mathbf{\Omega}) \mathbf{e}_r - \mathbf{e}_r \times \mathbf{v},
\end{equation}
the tidal torque can be written as

\begin{equation}
    \mathbf{N}_{Ti} = P_{to,i} \left[ \mathbf{\Omega}_i r_i - (\mathbf{r}_i \cdot \mathbf{\Omega}_i) \mathbf{e}_{ri} - \mathbf{e}_{ri} \times \mathbf{v}_i \right]
\end{equation}
\begin{equation}
    \mathbf{N}_{T\star} = P_{to,\star} \left[ \mathbf{\Omega}_\star ri - (\mathbf{r}_i \cdot \mathbf{\Omega}_\star) \mathbf{e}_{ri} - \mathbf{e}_{ri} \times \mathbf{v}_i \right]
\end{equation}

In heliocentric coordinates the spin evolution takes the form \citep{Bolmont2015}
\begin{equation}
    \frac{d}{dt} (I_\star \mathbf{\Omega}_\star) = - \sum_{j=1}^N \frac{m_\star}{m_\star + m_j} \mathbf{N}_{T\star}
\end{equation}
and
\begin{equation}
    \frac{d}{dt} (I_i \mathbf{\Omega}_i) = - \frac{m_\star}{m_\star + m_i} \mathbf{N}_{Ti},
\end{equation}
where $I$ is the moment of inertia.

We integrate the tidal spin evolution with the implicit midpoint method described in section \ref{sect:implicit} so that the total angular momentum of the system is conserved and we accurately model the evolution of the semi-major axis. Since every particle is affecting the spin of the central star, an additional parallel reduction step is necessary to sum up all contributions synchronously \footnote{
With a parallel reduction method, arrays of length $N$ can be summed up in log($N$) steps. Depending on the size $N$, this can be performed within a thread block, or must be distributed across different thread blocks and kernel calls.
}.

Note that our current implementation respects only the tidal effect from the central star to the other bodies, and vice versa. It does not include tidal effects between other bodies, such as tidal effects between multiple planets. This feature is already implemented in other codes (e.g. Posidonius \citep{Blanco-Cuaresma+Bolmont}), and we plan to eventually include it in a future version of GENGA.


\subsection{Test of the tidal forces}
\label{sect:tidalTest}
To test the implementation of the tidal forces, we repeat the calculations of the Mars-Phobos system described in \cite{Efroimsky2007} and \cite{Auclair-Desrotour+2014}. As initial values, we use the same parameters as given in Table 1 in \cite{Auclair-Desrotour+2014}. Additionally, we use an eccentricity of 0.0151 for Phobos. We compare our results against three secular evolution models. These are
\footnote{All three secular evolution models consider only the tidal bulge on the primary body, caused from the orbiting body, and not vice versa. A model which contains both directions is given in \cite{Correia2009}}

\begin{itemize}
    \item \citet[Equation 9 and 10]{Hut1981} \\
    This uses a constant $\tau$ value.
    
    \item \citet{Barnes2017}, constant $Q$ value\\
    (also called constant-phase-lag-model, CPL). \\It is similar to  \citet[Equation 30]{Efroimsky2007} (Kaula model).

    \item \citet{Barnes2017,Correia2009}, constant $\tau$ value\\
    (also called constant-time-lag-model, CTL). \\It is similar to  \citet[Equation 36]{Efroimsky2007} (Singer-Mignard model).

\end{itemize}

\begin{figure}
\includegraphics[width=1.0\linewidth]{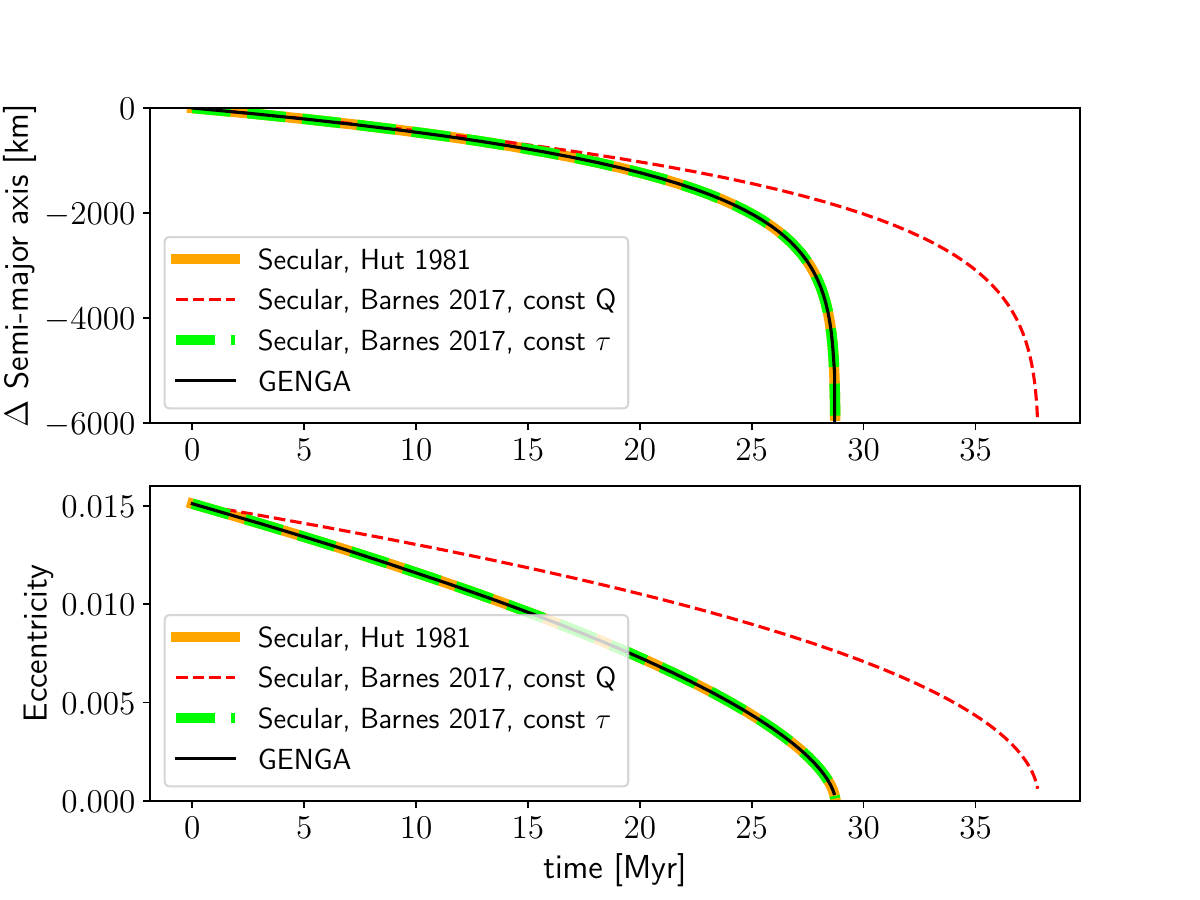}
\caption{Comparison of GENGA against secular evolution models on the example of the Mars-Phobos system. The models correspond to \cite{Hut1981} (orange solid line), \cite[const. $Q$ model]{Barnes2017} (red dashed line), and  \cite[const. $\tau$ model]{Barnes2017} (green dashed line). The evolution of the semi-major axis in GENGA agrees well with the constant $\tau$ scenario in the secular evolution models. The evolution of the eccentricity starts with the same slope, but then gets damped more rapidly in GENGA. This plot reproduces the results of Figure 2 from \cite{Efroimsky2007}.
}
\label{fig:PhobosTides} 
\end{figure}

The results are shown in Figure \ref{fig:PhobosTides}. The evolution of the semi-major axis and the eccentricity from GENGA agree well with the constant $\tau$ secular evolution models, as is expected because GENGA also uses a constant $\tau$ model. However, in GENGA the tidal model is not implemented as a secular evolution model. While we use a time step of 0.1 day in GENGA, the secular evolution models can use a time step of 100-1000 days.

\subsection{Rotational deformation}

When a viscous body is rotating, its shape is transformed into a symmetric oblate ellipsoid. The potential energy of a system containing two oblate bodies 0 and 1 is given in \citet[Equation 158]{Moyer1971}, \cite{Moyer2003} or \citep{Correia2011} as

\begin{equation}
    U = -\frac{Gm_0 m_1}{r} \left[ 1 - \sum_{i=0,1} \sum_{n=1}^{\infty} J_n \left( \frac{R_i}{r} \right)^n P_n(\hat{r} \cdot \hat{\Omega_i}) \right],
\end{equation}
with the mass $m$, the physical radius $R$, the distance between the bodies $r$, the rotational angular velocity $\Omega$ and the Legendre polynomial of degree $n$, $P_n(x)$. In GENGA we truncate the order $n$ to 2 \citep{Correia2011} and we compute the $J_2$ parameter purely via rotational deformation as \citep{Correia2011, Bolmont2015} 

\begin{equation}
J_{2,i} = k_{2f,i} \frac{\Omega^2_i R^3_i}{3G m_i}
\end{equation}
and
\begin{equation}
J_{2,\star} = k_{2f,\star} \frac{\Omega^2_\star R^3_\star}{3G m_\star},
\end{equation}
where $k_{2f,i}$ is the second potential Love number (fluid Love number) of body $i$.

We follow the description in \cite{Bolmont2015} and define the following quantities
\footnote{In \cite{Bolmont2015} $C_\star$ and $C_i$ are reversed.}:
\begin{equation}
C_{\star} = \frac{1}{2} G m_i m_\star J_{2,\star} R^2_\star
\end{equation}
and
\begin{equation}
C_{i} = \frac{1}{2} G m_i m_\star J_{2,i} R^2_i.
\end{equation}

With these definitions the force due to the rotational deformation is given as \citep{Bolmont2015}
\begin{eqnarray}
\mathbf{F_R} = \left\{ - \frac{3}{r^5_i} \left( C_\star + C_i\right) \right. \\ \nonumber
\left. + \frac{15}{r^7_i} \left[ C_\star \frac{(\mathbf{r}_i \cdot \mathbf{\Omega}_\star)^2}{\mathbf{\Omega}_\star^2} +  C_i \frac{(\mathbf{r}_i \cdot \mathbf{\Omega}_i)^2}{\mathbf{\Omega}_i^2} \right] \right\} \mathbf{r}_i \\ \nonumber
- \frac{6}{r^5_i} \left( C_\star \frac{\mathbf{r}_i \cdot \mathbf{\Omega}_\star}{\mathbf{\Omega}_\star^2} \mathbf{\Omega}_\star
+ C_i \frac{\mathbf{r}_i \cdot \mathbf{\Omega}_i}{\mathbf{\Omega}_i^2} \mathbf{\Omega}_i\right).
\end{eqnarray}

Similar to the tidal force, the rotational deformation generates a torque on the bodies.
This rotational deformation torque is given as \citep{Hut1981}
\begin{equation}
    \mathbf{N}_R = \mathbf{r} \times \mathbf{F}_{R\theta},
\end{equation}
where once again $\mathbf{F}_{R\theta}$ is the transverse component of the rotational deformation force. Following \cite{Bolmont2015} the torque can be written as

\begin{equation}
\mathbf{N}_{R\star} = - \frac{6}{r^5_i}  C_\star \frac{\mathbf{r}_i \cdot \mathbf{\Omega}_\star}{\mathbf{\Omega}_\star^2}  \left( \mathbf{r}_i \times \mathbf{\Omega}_\star \right)
\end{equation}
and
\begin{equation}
\mathbf{N}_{Ri} = - \frac{6}{r^5_i}  C_i \frac{\mathbf{r}_i \cdot \mathbf{\Omega}_i}{\mathbf{\Omega}_i^2}  \left( \mathbf{r}_i \times \mathbf{\Omega}_i \right).
\end{equation}
The rotational deformation torque is implemented the same way as the tidal torque described in section \ref{sect:tides}.

\section{Forces and models for small bodies}
\label{sect:smallBodies}
Simulating small bodies of the solar system (or other planetary systems) often requires the integration of a very high number of particles. There are more than 1 million asteroids known in the asteroid belt (https://ssd.jpl.nasa.gov/), and taking into account also the (unseen) smaller fragments would lead to immense numbers, exceeding current computational capabilities. Still, the presence of all these small bodies is important for the dynamics of meteoroids: collisions can lead to both break-up events and changes in meteoroid rotation rate. Together with the Yarkovsky effect and Poynting-Robertson drag, collisions between small bodies can influence the migration rate of asteroids and thus generate impact events on the Earth or other planets. An alternative to simulating such collisions with an $N$-body integrator is to include a probabilistic collision model, which uses an average probability that a given small body would collide with another small body \citep{Farinella1998}. Note that the presented probabilistic collision model is different from the real collision handling of GENGA and is applied only to test particles. Collisions between massive bodies or between a massive body and a test particle can still occur in the usual way (see Section \ref{sect:Collisions}). We do not include the YORP effect in this work, although it can also have an important contribution \citep{Rubincam2000, Vokrouhlicky2015}. While the Yarkovsky effect only alters the orbital elements of small radiated bodies, the YORP effect would also change the rotation rate and the obliquity of the body. Contrary to the Yarkovsky effect, the YORP effect depends on the shape of the object, which makes it harder to apply for real solar system asteroids.
In this section, we describe first the implementation of the Yarkovsky effect and the Poynting-Robertson drag, followed by a description of a model for collisional break-up events and rotation changes of small bodies. At the end of this section, we present an example simulation including these effects.

\subsection{Yarkovsky Effect}
\label{sect:Yarkovsky}
The Yarkovsky effect is caused by thermal radiation forces, acting on small rotating objects \citep{Oepik1951}. Typically, the Yarkovsky effect is important for asteroids or meteoroids with a diameter smaller than $\sim$40 km \citep{Bottke2006}. It causes a migration in the semi-major axis and is especially important to deliver material into mean-motion or secular resonances with the giant planets, where the affected material undergoes further dynamical effects such as eccentricity pumping (e.g. \citep{Broz2006}). 

The Yarkovsky effect emerges from the anisotropic re-emission of thermal radiation from the heated surface of the rotating object, which leads to an orbital acceleration. The effect can be split into two different types, the diurnal Yarkovsky effect and the seasonal Yarkovsky effect. The diurnal effect is the strongest when the spin axis is perpendicular to the orbital plane and can cause both inward migration and outward migration depending on the spin direction. The seasonal effect is the strongest when the spin axis of the object lies in the orbital plane, and leads to an inward orbital migration. The physical parameters and their values that go into computing the Yarkovsky effect are summarized in Table\, \ref{tab:Yarkovsky2}, and a more detailed description of the Yarkovsky effect is given in, e.g. \cite{Farinella1998,Vokrouhlicky1998, Vokrouhlicky1999, Vokrouhlicky2000, Bottke2000, Bottke2006}  or \cite{Broz2006}.

Following \cite{Vokrouhlicky1998}, we define the radiation force factor
\begin{equation}
    \Phi = \frac{\pi R^2 F}{m c},
\end{equation}
with the physical radius $R$, the mass $m$, the speed of light $c$ and the Solar radiation flux $F$ (in flux units) at the heliocentric distance $d$ given by
\begin{equation}
 F = \frac{S}{d^2}.
\end{equation}
Here $S$ is the Solar constant in W m$^{-2} $ and $d$ is the time averaged heliocentric distance in au, i.e.,
\begin{equation}
 d = a (1 + {\textstyle{\frac{1}{2}}}e^2),
\end{equation}
with $a$ the semi-major axis and $e$ the orbital eccentricity.

Following \cite{Vokrouhlicky1998} and \cite{Vokrouhlicky1999} further, we define the thermal inertia $\Gamma$ as
\begin{equation}
    \Gamma = \sqrt{\rho C K},
\end{equation}
which depends on the material density $\rho$, the specific heat capacity $C$ and the thermal conductivity $K$. We also define the thermal parameter
\begin{equation}
    \Theta_{\rm seasonal} = \frac{\Gamma \sqrt{n}}{\epsilon \sigma T_{\star}^3},
\end{equation}
where $\sigma$ is the Stefan-Boltzmann constant, $T_{\star}$ the subsolar surface temperature, and $n$ is the orbital mean motion. The subsolar surface temperature is set by
\begin{equation}
    T_{\star}^4 = \frac{(1-A)F}{\epsilon \sigma},
\end{equation}
with the Bond albedo $A$ and the thermal emissivity $\epsilon$.
Following the derivation in \cite[Appendix B]{Vokrouhlicky1999}, the acceleration of the seasonal Yarkovsky effect can then be written as
\begin{align} 
\label{eqn:YarkovskyS}
\mathbf{a}_{\rm seasonal} = \frac{4}{9}\frac{(1 - A) \Phi}{(1 + \lambda)} \times  \\ 
\sum_{k \ge 1} G_k\left[s_P \alpha _k \cos (knt + \delta_k) + s_Q \beta_k \sin(knt + \delta_k) \right] \mathbf{s}, \nonumber 
\end{align}
with the spin vector $\mathbf{s}$ and the  quantities $\lambda$, $\alpha$, $\beta$, $(G_k \cos \delta_k)$, $(G_k \sin \delta_k)$, $s_P$ and $s_Q$ as defined in \cite[Appendix B]{Vokrouhlicky1999}. In the current implementation in GENGA, we only use $k = 1$ in the summation of equation \ref{eqn:YarkovskyS}.

Similarly, we follow \cite{Vokrouhlicky2000} for the diurnal Yarkovsky effect and define
\begin{equation}
    \Theta_{\rm diurnal} = \frac{\Gamma \sqrt{\omega}}{\epsilon \sigma T_{\star}^3},
\end{equation}
with the rotational frequency $\omega$. The diurnal effect can then be written as
\begin{eqnarray} 
\mathbf{a}_{\rm diurnal} = \frac{4}{9}\frac{(1 - A) \Phi}{(1 + \lambda)}
G\left[\sin \delta + \cos \delta \mathbf{s} \times \right] \frac{\mathbf{r} \times \mathbf{s}}{r}.
\end{eqnarray}

The quantity $\delta$ is a thermal lag angle \citep{Vokrouhlicky1999}.
After having computed the total Yarkovsky acceleration
\[\mathbf{a}_Y = \mathbf{a}_{\rm seasonal} + \mathbf{a}_{\rm diurnal},
\]
the new velocity of the body is updated with a velocity kick of the time step $dt$ as
\[
\mathbf{v}^{n+ 1} = \mathbf{v}^{n} + \mathbf{a}_Y dt.
\]

\subsection{Orbit-averaged Yarkovsky effect}
As an alternative to calculating the acceleration of the Yarkovsky effect at every time step, an orbit-averaged drift in the semi-major axis can be applied.

For circular orbits and by linearizing the heat conductivity, the orbit-averaged change in the semi-major axis $a$ can be written following \cite{Vokrouhlicky1999} and \cite{Vokrouhlicky2000} as
\begin{equation}
\label{eqn:dadt_diurnal}
\left(\frac{da}{dt} \right)_{\rm diurnal} = - \frac{8}{9} \frac{(1-A)\Phi}{n} \frac{G \sin \delta}{1 + \lambda} \cos \gamma
\end{equation}
and
\begin{equation}
\label{eqn:dadt_seasonal}
\left(\frac{da}{dt} \right)_{\rm seasonal} = \frac{4}{9} \frac{(1-A)\Phi}{n} \sum_{k \ge 1} \frac{G_k \sin \delta_k}{k} \chi_k \bar{\chi_k},
\end{equation}
where $\gamma$ is the spin axis obliquity, and the quantities $G$, $\delta$, $\lambda$ and $\chi $ are defined in \cite{Vokrouhlicky1999} and \cite{Vokrouhlicky2000}.
Similar formulations are derived in \cite{Farinella1998}, \cite{Vokrouhlicky1999}. 
A more precise model of the Yarkovsky effect is described in \cite{Vokrouhlicky2000} or \cite{VokrouhlickFarinella1998}, and involves the numerical integration of the heat diffusion equation. For simplicity, we do not use this model in the current implementation of GENGA. The user can enable the direct Yarkovky effect or the orbit-averaged Yarkovsky effect in the GENGA parameter file.
Applying the averaged equations \ref{eqn:dadt_diurnal} and \ref{eqn:dadt_seasonal} requires a conversion between Cartesian coordinates to Keplerian elements and back, which causes computational overhead.

\subsection{Test of the Yarkovsky effect}
\label{sect:YarkovkyTest}
\begin{figure}
\includegraphics[width=1.0\linewidth]{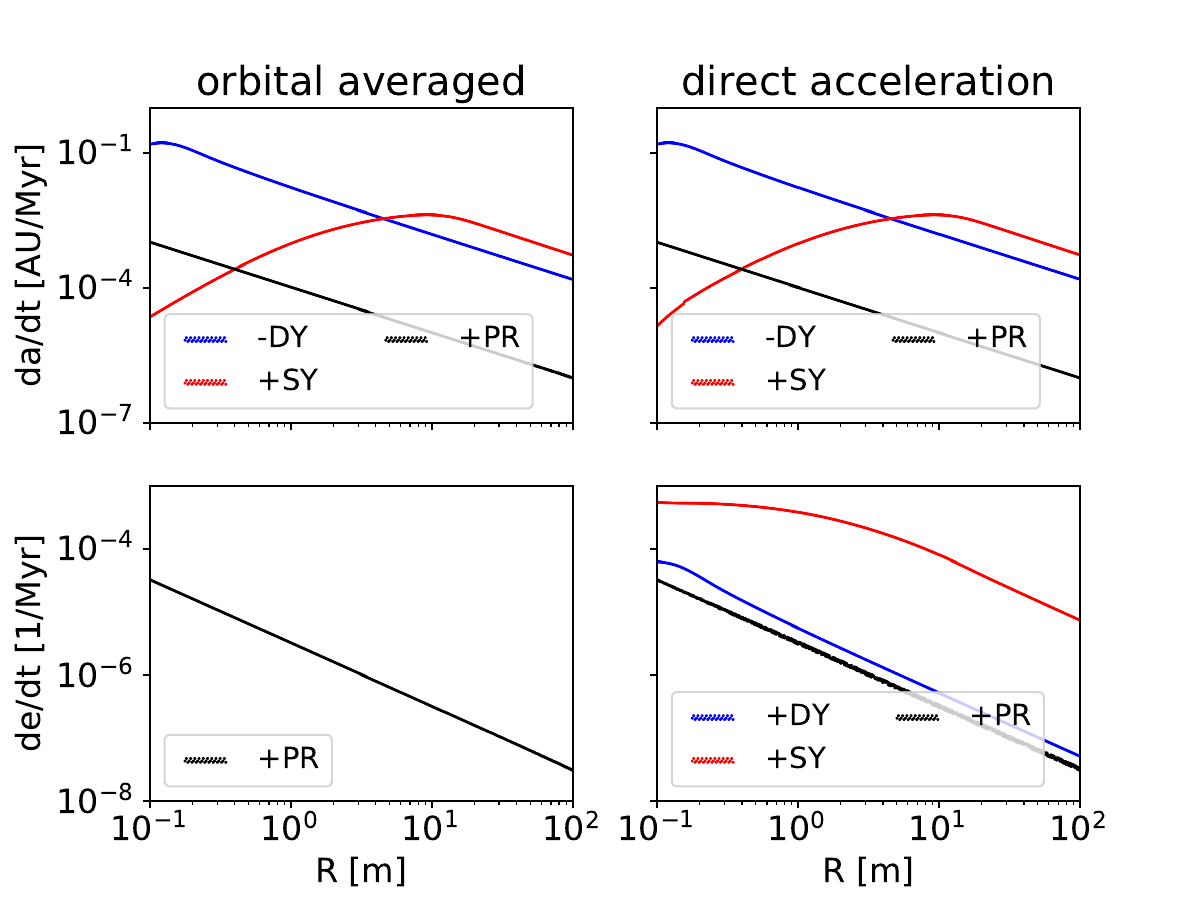}
\caption{Test of the seasonal and diurnal Yarkovsky effect and the Poynting-Robertson drag for the values of Table \ref{tab:Yarkovsky2} (For the Poynting-Robertson drag, we use e = 0.05 instead of 0). Shown are both methods, the orbital-averaged drift method $da/dt$ ($de/dt$) (left panels) and the direct acceleration $\mathbf{a_Y}$, $\mathbf{a_{PR}}$ (right panels). Note that the orbital-averaged drift method for the Yarkovsky effect only affects the semi major axis and not the eccentricity. 
This plot reproduces the result from \citep[Figure 4]{Bottke2000} for the Yarkovsky effect.}
\label{fig:Yarkovsky} 
\end{figure}

We test the Yarkovsky effect with the same initial conditions as given in \cite[Section 4]{Farinella1998} and \cite[Section 4.1]{Bottke2000}, which are summarized in Table \ref{tab:Yarkovsky2}. We generate 1000 test particles with a physical radius between 0.1 m and 100 m and integrate them for 10,000 yr with both methods: the direct and the orbital-averaged method. The direct acceleration method is simpler to apply into the $N$-body integrator and is in practice also faster than the orbital-averaged drift method because it does not require the conversion to orbital elements and back again. Therefore we recommend to use the direct acceleration method in simulations that include the Yarkovsky effect. 
The results are shown in Figure \ref{fig:Yarkovsky} reproduce the results from \cite[Figure 4]{Bottke2000}. Note that there is a factor of 2 difference between the results of \cite[Figure 4]{Bottke2000} and \cite[Figure 1]{Farinella1998}.

Another example including the Yarkovsky effect is presented in section \ref{sect:Hebe}.

\begin{table}[ht]
\centering
\caption{Parameters and initial conditions for the Yarkovsky effect test}
\label{tab:Yarkovsky2}
\begin{tabular}{l l l }
Semi-major axis & $a$ & 2 au \\
Eccentricity & $e$ & 0 \\
Inclination & $i$ & $0^\circ$ \\
Material density & $\rho$ & 3500.0 kg/m$^3$ \\
Physical radius & $R$  & 0.1 - 100 m \\
Rotation frequency & $\omega$ & 5 h \\
Obliquity & $\gamma_{\text{Seasonal}}$ & 90$^\circ$ \\
Obliquity & $\gamma_{\text{Diurnal}}$ & 0\\
Thermal conductivity & $K$ & 2.65 W m$^{-1}$ K$^{-1}$\\
Specific heat capacity & $C$ & 680 J kg$^{-1}$ K$^{-1}$\\
Thermal emissivity & $\epsilon$ & 1.0 \\
Asteroid's Bond albedo & $A$ & 0.0\\
Solar constant & $S$ & 1367 W/m$^2$.
\end{tabular}
\end{table}

\subsection{Poynting-Robertson drag}
\label{sect:PR}
When a body is so small that its orbital momentum becomes comparable to the total momentum caused by solar photons (typically below the centimeter to millimeter range) it encounters, it becomes subject to Poynting-Robertson drag, for which many different derivations are known, e.g. \cite{Robertson1937,Burns1979,Burns2004}.

Following \cite{Burns1979}, the change in the semi-major axis $a$ and the eccentricity $e$ due to this force can be written as

\begin{equation}
\label{PR1}
\frac{da}{dt} = -\frac{\eta}{a}Q_{pr} \frac{(2 + 3e^2)}{(1 - e^2)^{3/2}}
\end{equation}
and
\begin{equation}
\label{PR2}
\frac{de}{dt} = - \frac{5}{2}\frac{\eta}{a^2}Q_{pr} \frac{e}{(1 - e^2)^{1/2}},
\end{equation}
with 
\begin{equation}
\eta = \frac{S_0r_0^2A}{mc^2}
\end{equation}
where $A$ is the cross section of the particle, $S_0$ is the Solar constant at 1 au, and $r_0$ = 1 au. The radiation pressure coefficient $Q_{pr}$ is assumed to be 1 for purely absorbing particles.

Implementing the equations \ref{PR1} and \ref{PR2} in an $N$-body integrator requires the transformation from Cartesian coordinates to Keplerian elements and back to Cartesian coordinates, after the terms \ref{PR1} and \ref{PR2} have been applied. These conversations would require a substantial amount of the total computing time. A more efficient approach is to apply the velocity change directly as \citep{Burns1979}

\begin{equation}
\label{eq:PR}
\mathbf{a}_{\rm PR} = \frac{d \mathbf{v}}{dt} = \frac{\eta c}{r^2} Q_{pr} \left[ \left(1 - \frac{\dot{r}}{c} \right) \hat{r} - \frac{\mathbf{v}}{c} \right].
\end{equation}

Since the acceleration in equation \ref{eq:PR} depends on the velocity, we use a few implicit midpoint iterations to perform the velocity kick operation in a symplectic way. Typically, less than three iterations in the implicit midpoint method are needed to converge to machine precision.

\subsection{Test of the Poynting-Roberton drag}
We repeat the same test as in section \ref{sect:YarkovkyTest} for the Poynting-Robertson drag, with the exception that here we use an eccentricity of 0.05 instead of 0 the get non-zero values for $de/dt$. The results are shown in Figure \ref{fig:Yarkovsky} together with the Yarkovky effect, and show that the Poynting-Roberson drag is important only for small particles with a radius $<$ 1 m.

\begin{figure}
\includegraphics[width=1.0\linewidth]{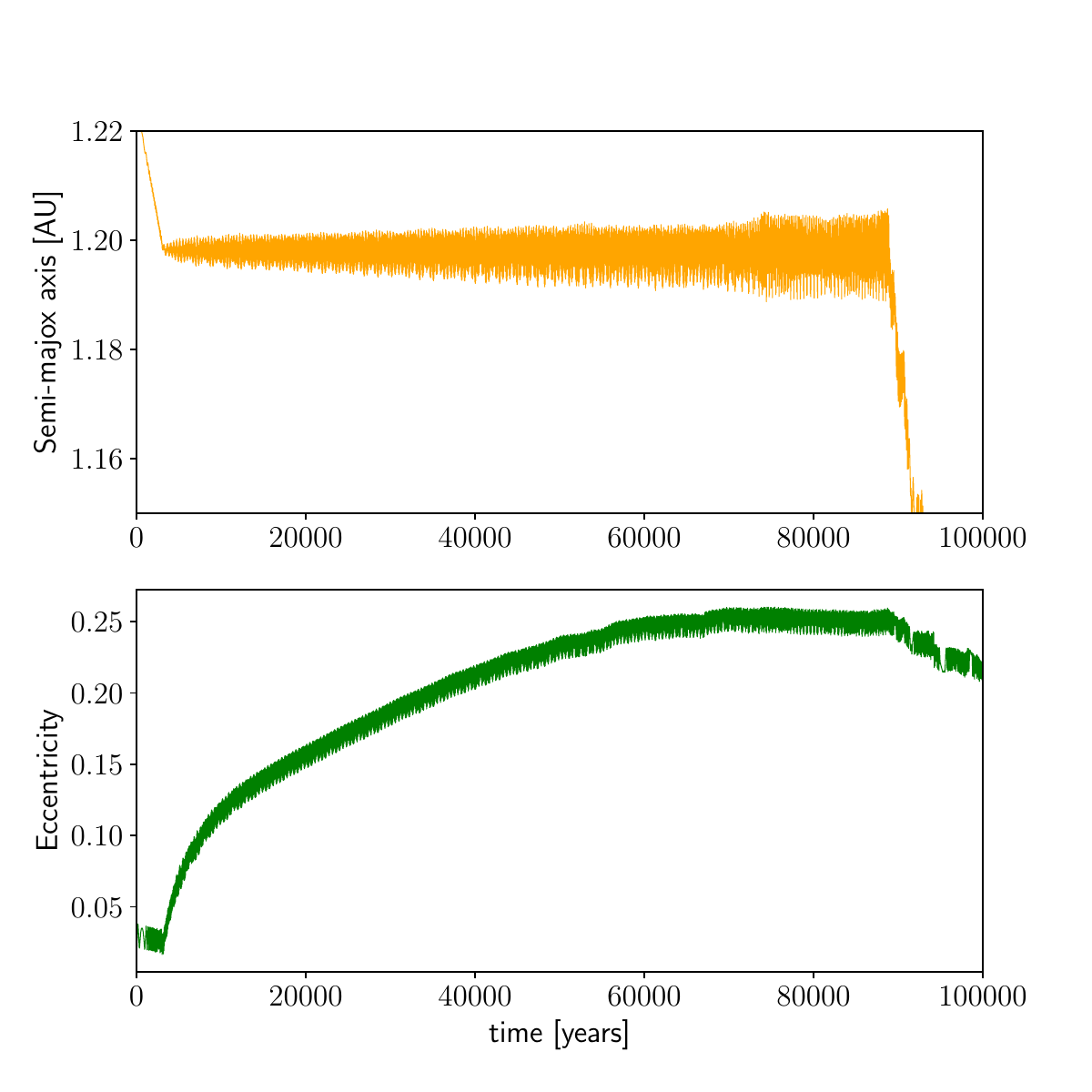}
\caption{Orbital evolution of a 28 $\mu$m sized dust particle under the influence of Poynting-Roberson drag and the presence of the solar system planets. The dust particle drifts toward the Sun and gets trapped in a resonance condition around the Earth. After 90,000 yr, the eccentricity gets large enough that the dust particle can leave the orbital trap again.}
\label{fig:PR} 
\end{figure}

For another test of the Poynting-Robertson drag we follow
\cite{Kortenkamp2013}, who uses numerical simulations to follow the orbital evolution of dust particles in the solar system. The Poynting-Robertson drag causes the semi-major axis to decay, and the particle drifts toward the Sun until it eventually reaches a resonance condition with a planet. Dust particles can then be trapped in this resonance condition for many tens of thousands of years. Similar to \cite{Kortenkamp2013}, we analyze the orbital evolution of a 28 $\mu$m sized dust particle in the presence of all plants of the solar system. In Figure \ref{fig:PR} the evolution of the semi-major axis and of the eccentricity of a dust particle are depicted. After 3000 yr, the dust particle gets trapped in orbital resonance with the Earth, specifically the 2:3, and the eccentricity begins to increase until it reaches a value of $\sim$0.25 and the particle leaves the resonance condition again after 90,000 yr.  

\subsection{Collisional break-up model}
\label{Collision}

The asteroid belt of the solar system contains a very large number of objects. This means that particles within the asteroid belt will eventually collide with some others. It is computationally not feasible to include all asteroids in every simulation so that it is desirable to describe the asteroid belt by a probabilistic collision model. Such a model uses the probability that an object would collide with another object in the asteroid belt and uses random numbers to generate virtual collision events. These events can either lead to a reset of the rotation rate of a given object or destroy the object and replace it with new generated fragment particles. In the following, we describe the details of the collisional break-up model, which generates new fragment particles on the fly, and the rotation reset model.
\textit{We emphasize that the collisional break-up model and the rotation-rate reset model are tailored to the asteroid belt between Mars and Jupiter.}

Following \cite{Farinella1998}, we use a collisional lifetime of
\begin{equation}
\tau_{col} = 20 \sqrt{\frac{R}{1\,\text{m}}}\: \text{Myr},
\end{equation}
which is equivalent to a break-up probability of  $1/\tau_{col}$ per year. When a small body breaks up, its mass is replaced by a series of fragments with radii between the radius of the original object $R_o$, and  $R_m$ = 0.01 m. The minimum radius $R_m$ is a compromise between run time and accuracy. Having a smaller value for $R_m$ would generate much more particles, which slows down the integration speed. Setting it to a larger value would generate less particles. The value of $R_m$ can be set as a user parameter. 
We use the following distribution to generate the radii of the new fragments $R_f$, which is given by

\begin{equation}
R_{f} = \left[ \left(R_0^{(n + 1)} - R_m^{(n + 1)} \right) \cdot y +  R_m^{(n + 1)} \right]^{1/(n + 1)},
\end{equation}
with $n$ = -3/2 and a uniform generated random number $y \in (0,1)$. The mass of each generated fragment is subtracted from the remaining mass of the original body. The first fragment to exceed the remaining mass gets the remaining mass.

For the velocity distribution of the fragments, we use a scaling value
\begin{equation}
u = y \cdot m_f^{-1/6},
\end{equation}
with the mass of the fragment $m_f$ and a random number $y \in (0.8, 1.2)$ \citep{nakamura1991}.
A constant velocity budget of 31 m s$^{-1}$ (based on $V_{\rm budget}$ = (0.3/$N$)$^{0.5}$ $\times V_{\rm impactor}$, and $N$ = 8000; see \cite{wiegert2015} for details) is distributed to the fragments in proportion to the scaling value $u$. A randomly distributed velocity vector with the given length sets the new orbit of the particle. Typical values of these created fragment velocities are of the order of $\sim0.1-1$ m s$^{-1}$.

In GENGA, fragments with a radius smaller than $R_\text{remove}$ = 0.01 m are removed to prevent having too many particles in the simulation. The value of $R_\text{remove}$ can be set as a user parameter. The spin axis direction is generated randomly following a two-dimensional bivariate normal distribution over the solid angle.
For the rotation rate, we take
\begin{equation}
\omega_f = {\rm random} \left( \frac{1}{36R_f}, \frac{1}{R_f}\right),
\end{equation} which reproduces the range of rotation rates typically observed for Near Earth Asteroids in the meter \citep{beechbrown2000} to kilometer \citep{Farinella1998} size range.

\subsection{Collisional reset model of rotation rate and obliquity}
\label{CollisionalReset}

For each object at each time step, we calculate the probability that its rotation was reset during the last time step by a virtual collision with another particle as a function of its radius $R$ and previous rotation rate. The rotation will be changed if the impactor has an angular momentum at the collision time similar to the rotational angular momentum of the target object, such that the radius of the projectile is given as \citep{Farinella1998}
\begin{equation}
r_{\rm projectile} = \left( \frac{2 \sqrt{2} R \omega \rho_{\rm target} }{5 \rho_{\rm projectile} V}\right)^{1/3} R
\end{equation}
which, assuming $\rho_{\rm target}= \rho_{\rm projecile}$, simplifies to
\begin{equation}
\label{eq:rpro}
r_{\rm projectile} = \left( \frac{2 \sqrt{2} \omega}{5 V}\right)^{1/3} R^{4/3}
\end{equation}
Here $V$ is the typical collisional velocity in the asteroid belt ($\sim 5$ km s$^{-1}$). The probability of a rotation-rate reset through collision with a large enough projectile is given by

\begin{equation}
\label{eq:collProb}
\frac{1}{\tau_{\rm rot}} = 3\times 10^5 P_i R^2 r_{\rm projectile}^{-5/2} \,{\rm s}^{-1}
\end{equation}
with $P_i$ being the intrinsic collisional probability for the asteroid belt, $2.85\times 10^{-18}\, \text{km}^2 \text{yr}^{-1}$ \citep{Farinella1998}. Using this number, and replacing $r_{\rm projectile}$ in equation \ref{eq:collProb} with the expression in equation \ref{eq:rpro} results in a final probability of

\begin{equation}
\frac{1}{\tau_{\rm rot}} = 10^{-18} R^{-4/3} \left[ \frac{2 \sqrt{2} \omega}{5 V}\right]^{-5/6} \,{\rm s}^{-1}
\end{equation}
where all values are in SI units. If the rotation rate is reset, then both the obliquity and the new rotation rate are randomized as described in section \ref{Collision}.

\subsection{Example simulation of debris particles}
\label{sect:Hebe}
\begin{figure*}[ht]  
\includegraphics[width=1.0\linewidth]{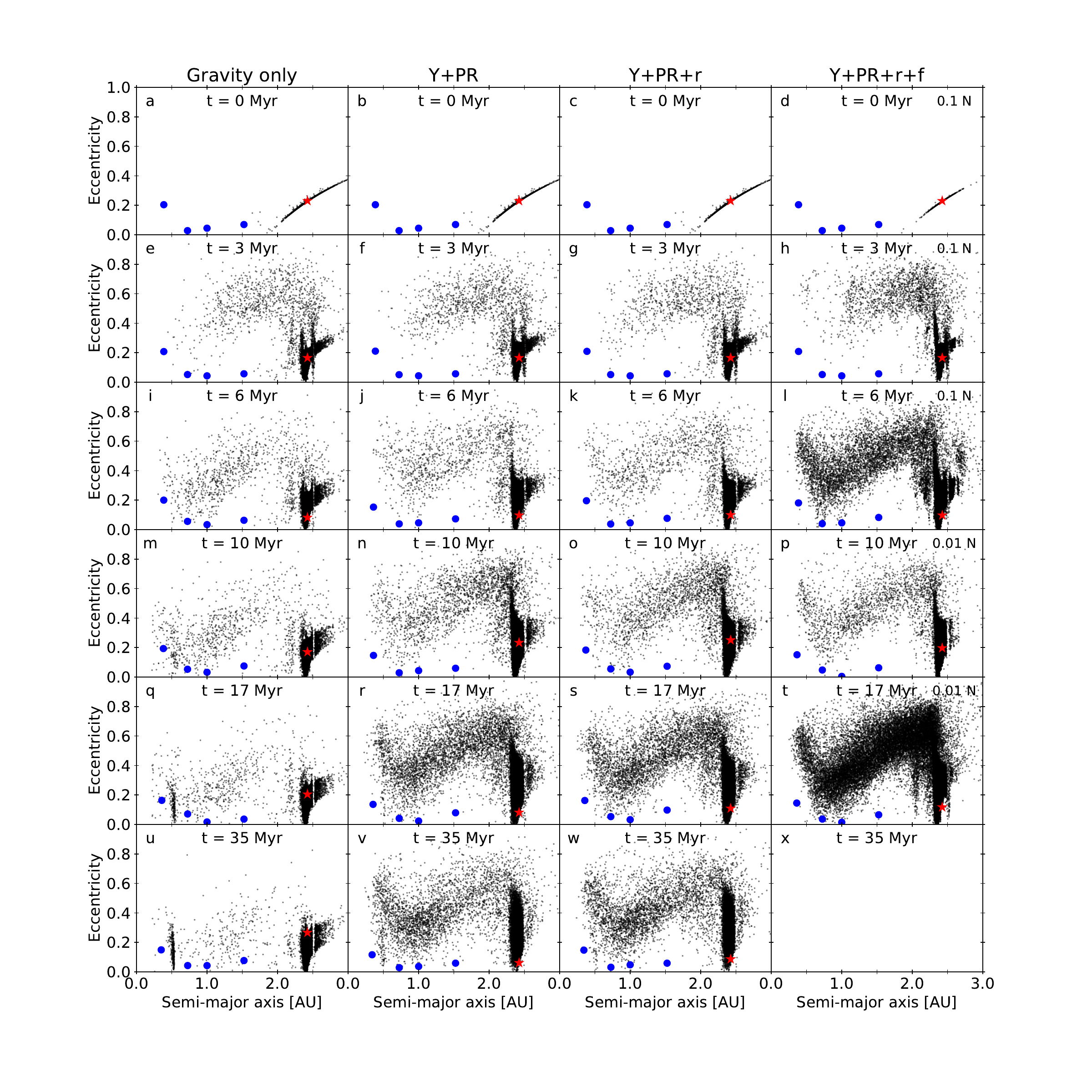}
  \caption{Evolution of ejecta material of an asteroid collision. The fragment particles migrate via the Yarkovksy effect and Poynting-Robertson drag into resonance conditions with Jupiter, where the eccentricities of the particles get increased. Later, the eccentricities get damped from the inner planets and finally they can reach Earth-crossing orbits and even hit the surface of the Earth.
  The left column shows the evolution of the fragment particles with only the gravitational force included, the second column includes the Yarkovsky effect and Poynting-Robertson grad (Y+PR). The third column includes the rotational reset model (Y+PR+r), and the fourth column includes the fragmentation model (Y+PR+r+f). The simulation in the fourth column starts with 10 times less particles as the others (indicated with '0.1 $N$'), and after 6 Myr, the number of particles is again reduced manually by a factor of 10 (indicated with '0.01 $N$') to limit the total number of particles.
  The blue dots show the positions of the inner planets and the red star indicates the position of the 6 Hebe parent body. The last sub-panel of the Y+PR+r+f simulation (sub-panel x) is empty because this simulation  only reached 17 Myr.
  }
  \label{fig:Hebe}
\end{figure*}

To test the new functions and forces for small bodies we simulate the transport of ejecta material from the asteroid 6 Hebe to the Earth, in a similar fashion to that described in \cite{Bottke2000} and \cite{Broz2006}. The asteroid 6 Hebe is thought to be a parent body of the H chondrites \citep{Gaffey1993}, one of the most abundant types of meteorites falling to Earth today. It is located close to mean-motion resonances with Jupiter (mainly the 7:2 and the 3:1 mean-motion resonances), which makes the transport of small material very efficient. The ejecta material migrates via the Yarkovsky effect into the resonance, where the eccentricity is increased and the particles can reach an Earth-crossing orbit and some particles will hit the surface of the Earth or also other inner planets \citep{Bottke2000}. 
As initial conditions of our simulation, we use the coordinates of the solar system planets and the asteroid 6 Hebe 50 Myr ago, which we obtain from a backward integration of today's positions and velocities. For the backward integration, we include the solar system's planets and Pluto, following the idea of \cite{Bottke2000}, where the planets Venus to Neptune are included. To obtain a more precise integration, also the largest asteroids of the solar system could be included. We note that also the precise time of the break-up event can influence the dynamics of the ejecta particles. Following \cite{Bottke2000}, we chose 50 Myr. At the location of 6 Hebe, we generate 39,574 test particles with random spins and radii following a power-law distribution with slope $n=-3/2$ between $r_0=0.01\,$m and $r_1 = 50\,$m. From there, we simulate the system forward in time with and keep track of all encounter and collision events with the planets.
In Figure \ref{fig:Hebe} are shown four simulations, all starting with the same initial conditions. The first simulation (left column) contains only gravity and no additional forces or models. Since the ejecta particles have an initial velocity, some of them are located already in resonant conditions with Jupiter. There, their eccentricities are increased and the particles migrate toward the inner solar system, where the inner planets damp the eccentricities again and the particles can reach Earth-crossing orbits or even collide with a planet. After about 10 Myr, the bulk of the ejecta particles has reached the inner solar system. 

The second column of Figure \ref{fig:Hebe} includes the Yarkovsky effect and Poynting-Robertson drag. With these two effects, debris particles can constantly drift into the resonances with Jupiter and lead to many more particles reaching the inner solar system over a protracted time period.

The third column of Figure \ref{fig:Hebe} adds the rotational reset model, described in section \ref{CollisionalReset}. The result looks very similar to the second column. Differences would only occur at a later time, where the rotation reset model leads to a random walk in the semi-major axis of the debris particles. 

The fourth column of Figure \ref{fig:Hebe} includes also the fragmentation model, described in section \ref{Collision}. Since the number of particles grows exponentially with this model, we start with only 10\% of the particles as initial conditions. In order to speed up the calculation, we stop the simulation after 6 Myr manually, remove 90\% of the particles (i.e. we reduce the number of particles from 126,564 to 12,656) and restart the simulation again from there. Additionally, the fragmentation model constantly removes all new generated particles with a radius of less than 0.01\,m. The fragmentation model leads to many more smaller particles, which can increase the migration rate of the particles into the resonances but also increases the computation time dramatically.

\begin{figure}[ht]  
\includegraphics[width=1.0\linewidth]{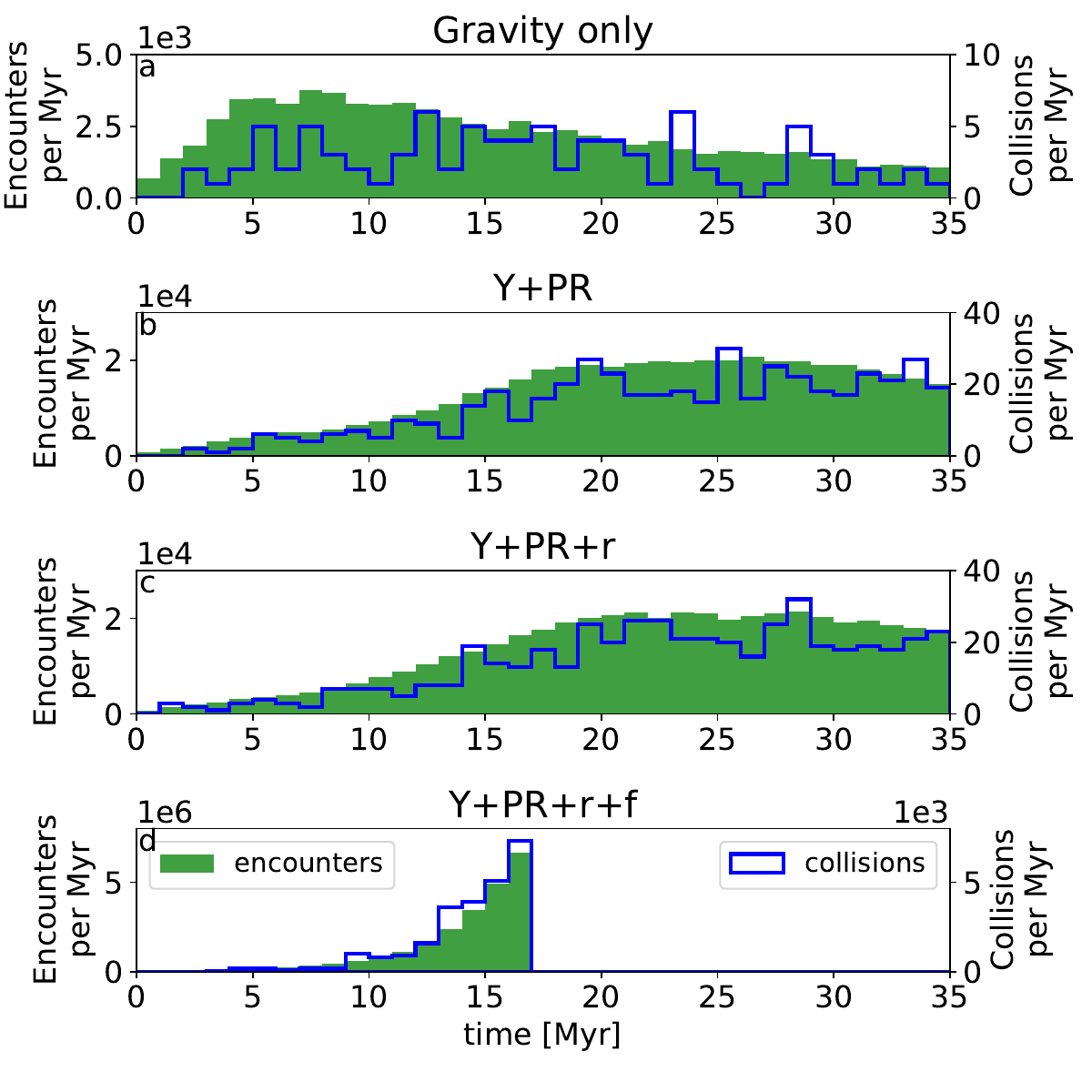}
  \caption{Close-encounter and collision events of the ejecta material with the Earth for the same simulations shown in figure \ref{fig:Hebe}. Close encounters in this Figure are encounter events with a closest distance to the Earth of less than 50 Earth radii.
  The data of the Y+PR+r+f simulation (panel d) is scaled up to the original number of initial particles.}
  \label{fig:HebeEnc}
\end{figure}

In Figure \ref{fig:HebeEnc} we show all encounter and collision events with the Earth. The aim of this figure is to display the different collision rates and collision timeline for the different forces that are applied.  The green bars display encounters with the Earth within 50 Earth radii, while the blue histograms show collisions with the Earth. It is clear that, in the gravity-only simulation, the number of encounters peaks between 5 and 10 Myr and then slowly declines. In the cases with Y+PR and Y+PR+r the number of encounters increases and remains roughly steady between 20 and 35 Myr. In the bottom panel the number of encounters and collisions increases rapidly until 17 Myr, after which the simulation is stopped.


\section{Improvements in the close-encounter integration and high-resolution simulations}
\label{sect:closeEncounters}
In the following section, we describe the updates to the close-encounter integration routines. First, it includes new parallelization schemes for the Bulirsch-Stoer kernels, which allow us to integrate much larger close-encounter groups. Second, we introduce a hierarchical method that uses multiple changeover functions, because it is able to integrate large close-encounter groups much more efficiently. We demonstrate how this new method can increase the computational speed significantly. 

\subsection{The Bulirsch-Stoer direct integration kernels}
\label{sect:BS}
in GENGA, close-encounters are integrated with a Bulirsch-Stoer method. All particles in a given close-encounter group need to be integrated synchronously. To achieve a good performance, it is necessary to always use the fastest available memory on the GPU\footnote{GPUs contain different memory types with different sizes and access speeds. The fastest and most limited are the local registers, followed by the shared memory. The largest and slowest memory type is the global memory. Differences in access times between these memory types can easily be a factor of 100 or more.
\\
The CUDA programming model is organized in threads, warps, thread blocks, and grids. The smallest units are the threads; they can be executed in parallel by the CUDA cores. A collection of 32 (or 64) consecutive threads is called a warp. All threads in a warp must perform the same operation. They can communicate via warp shuffle functions with registers or via shared memory. A thread block can contain up to 1024 threads. Threads in a thread block can communicate via shared memory, they cannot communicate with threads in other thread blocks. Communication between thread blocks is only possible via the CPU and different kernel calls and by using global memory.}.
While small close-encounter groups can be integrated within a single thread block and benefit from fast shared memory, larger groups must be distributed across multiple thread blocks and are more affected by a slower global memory access and kernel overhead time.
Depending on the close-encounter group size $n$, different parallelization methods are needed. In the following, we describe four different methods to cover the range from $n=2$ to $n \gg 1000$ and summarize the parallelization parameters used in Table \ref{tab:BS}. GENGA sorts all close-encounter pairs into independent groups, which can be integrated in parallel. Different groups of the same size can be easily launched in parallel by using multiple thread blocks. To launch different versions of the kernels, different CUDA streams can be used 
\footnote{A CUDA stream is a sequence of GPU operations, e.g. kernel calls or memory transfers. Different CUDA streams can be executed concurrently in parallel}. 
Table \ref{tab:BS} indicates the stream index, the number of threads per thread block and the number of thread blocks per close-encounter group of the different Bulirsch-Stoer kernel implementations.

\begin{table}[ht]
\caption{Parameters for the close-encounter integration parallelization.}
\label{tab:BS}
\begin{tabular}{c | c | c | c | c}
 Group & Kernel & Stream & Threads    & Thread \\
 Size  & Name   & Index  & per Block & Blocks \\
 \hline
 2      & BS-B & 0 & 4 &  1 \\
 3-4    & BS-B & 1 & 16 & 1 \\
 5-8    & BS-B & 2 & 64 & 1 \\
 9-16   & BS-B & 3 & 256 & 1 \\
 17-32  & BS-B & 4 & 256 & 1 \\
 33-64  & BS-A & 5 & 64 & 1 \\
 65-128 & BS-A & 6 & 128 & 1 \\
 129-256 & BS-A & 7 & 256 & 1 \\
 255-512 & BS-A or BS-M & 8 & 256 & 1 or 4 \\
 513-1024 & BS-M & 9 & 256 & 4 \\
 1025-2048 & BS-M & 10 & 256 & 8 \\
 ... & BS-M & ... & 256 & ... 
 
\end{tabular}
\end{table}
The different close-encounter kernels are BS-B, BS-A and BS-M, which are described in detail below. The BS-M kernels cannot use more than 256 threads per thread block, because of a hardware limitation in the number of registers.

\subsubsection{The BS-B kernel}
\cite{GrimmStadel2014} describes an implementation of the Bulirsch-Stoer method (BS-B) by using shared memory and a full $n^2$ parallelization, where $n$ is the size of the close-encounter group. This implementation is very efficient for small values of $n$, but it is not applicable for large $n$ because of limitations of the shared memory size on the GPUs and the number of threads per thread block. Due to hardware limitations, we can use this kernel only for groups up to a size of 32 bodies. The BS-B kernel is still identical to the description in \cite{GrimmStadel2014}.

\subsubsection{The BS-A kernel}
The BS-A kernel does not use an $n^2$ neighbor accessing scheme, but it uses a list of all close-encounter partners, and all involved particles iterate through that list. Since not all particles must have the same number of close-encounter pairs, this implementation can lead to thread divergences
\footnote{A thread divergence happens if some threads in a warp should execute a different operation than others. Since this is not possible, the two operations are serialized and performed one after the other. In that case, the number of active threads in the warp is reduced, which is called a lower occupancy. The occupancy depends on the number of bodies and the initial conditions. For a simulation with 8192 particles, we measured an occupancy of about 75\%.},
which affects the occupancy of the kernel. Reading and iterating the close-encounter pairs list requires additional computing time. The BS-A kernel is less efficient than the BS-B kernel for small values of $n$, and since the BS-B kernel is not scalable to $n>32$, it is the best alternative for groups sizes up to 256 due to limitations on shared GPU memory. Even larger close-encounter groups would need too much shared memory and another routine is required. 

\subsubsection{The BS-M function}
For $n>256$, the integration cannot be parallelized further within only a single thread block, but instead the close-encounter groups must be distributed over multiple thread blocks. This also requires that the sub-steps of the Bulirsch-Stoer integration are controlled from the host (i.e. the CPU) and therefore this involves many different kernel calls and communication between CPU and GPU over the motherboard and access to global memory. It is clear that this method suffers from kernel overhead time and global memory access which slow the computation down due to communication bottlenecks. On the positive side, this method can be applied to all group sizes. 

\subsection{Close-encounter performance test}
\begin{figure}
\includegraphics[width=1.0\linewidth]{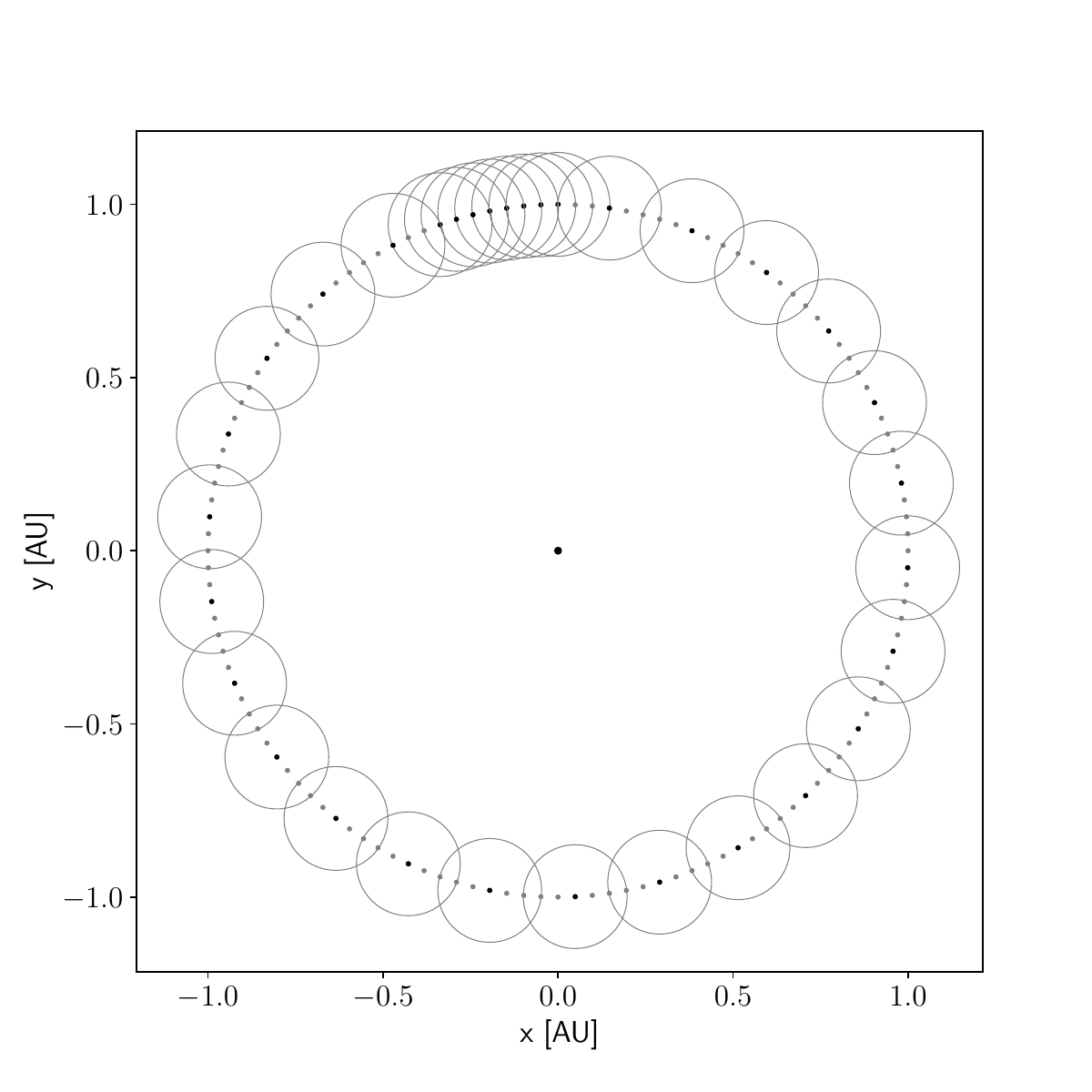}
\caption{Initial conditions for the close-encounter performance test. $N$ equal-mass particles are regularly positioned in a ring around a central mass. The critical radii of the particles is chosen such that every particle is in a close-encounter with eight other neighboring particles. Since the critical radii overlap, all particles are grouped together into a single close-encounter group. The figure shows an example with $N$ = 128. The circles show the critical radii of the particles, but for clarity of the figure, we do not show them for all particles.}
\label{fig:ring} 
\end{figure}

\begin{figure}
\includegraphics[width=1.0\linewidth]{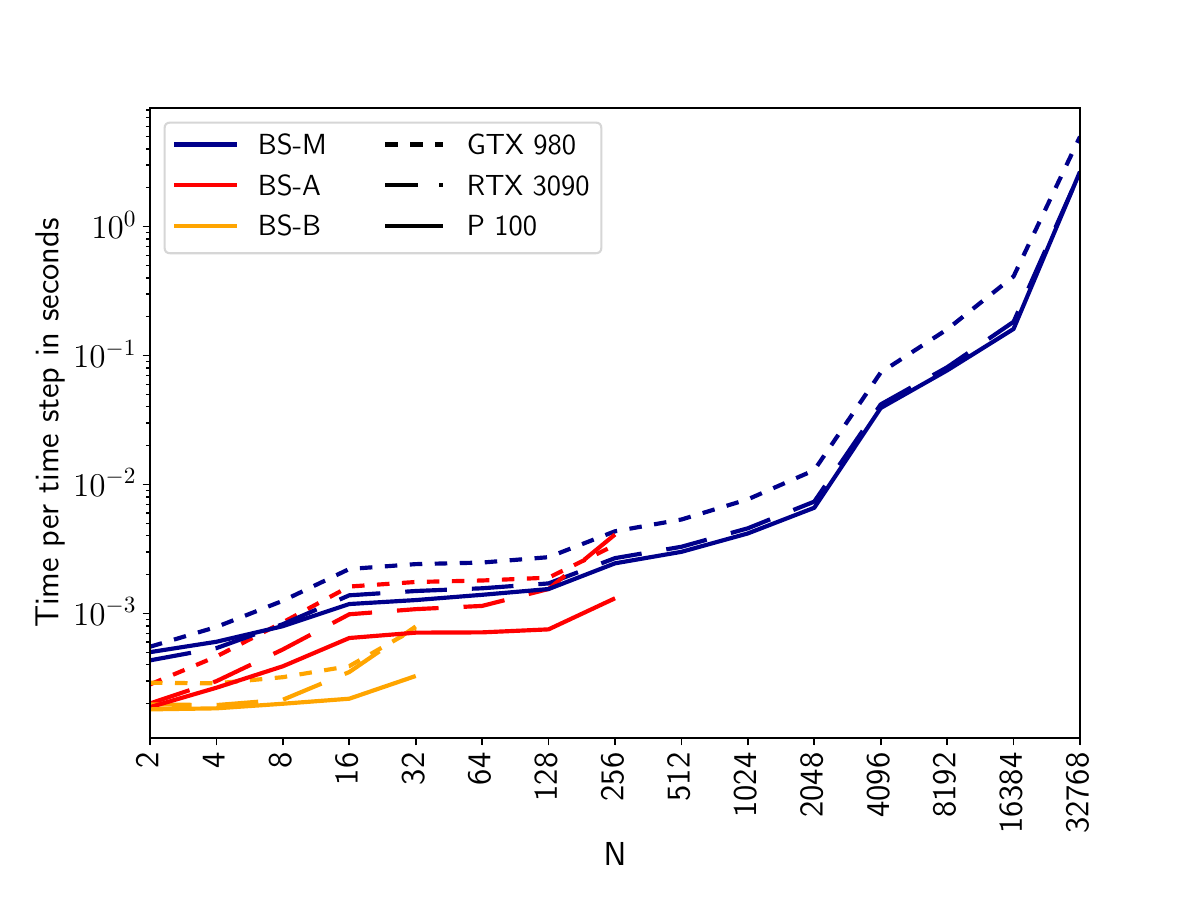}
\caption{Performance of different parallelization implementations of the Bulirsch-Stoer method used for the close-encounter integration. The used initial conditions for this speed test are shown in Figure \ref{fig:ring}. The BS-B method benefits from the fast data communication through shared memory, but it is not scalable to more than 32 particles. For larger group sizes, more and more global memory must be used, which leads to a performance penalty. The steep ascend from $N$=2 to $N$=16 for the BS-A and BS-M methods comes from the fact that there are not enough neighbor particles available for eight close-encounter partners. Shown is the data for three different GPU types, a GTX 980, a RTX 3090 and a Tesla P100.}
\label{fig:BSTime} 
\end{figure}

To test the performance of the different implementations as a function of group sizes, we position $N$ particles regularly spaced on a ring around a central mass as shown in Figure \ref{fig:ring}. We set the critical radii of the particles in a way that every particle is in a close-encounter with eight neighboring particles (for $N<16$ we use all available neighbors). In this way, all $N$ particles are grouped together into a single close-encounter group. The Figure \ref{fig:BSTime} shows the execution time per time step of the different methods, depending on the number of bodies $N$. The used initial conditions are very synthetic and in real simulations, there are certainly situations where the performance of the different implementations can differ, but it demonstrates well the importance of fast memory usage and kernel overhead. This test shows also that integrating large close-encounter groups is the performance bottleneck in simulations with many particles. In practice, large close-encounter groups $(\gtrsim 128)$ should be avoided if possible; in section \ref{sect:SLevels}, we present a way to achieve that. In the latest version of GENGA, we use the BS-B, BS-A and BS-M methods for close-encounter group sizes, as indicated in table \ref{tab:BS}.

\subsection{Higher level changeover functions}
\label{sect:SLevels}
\begin{figure}
\includegraphics[width=1.0\linewidth]{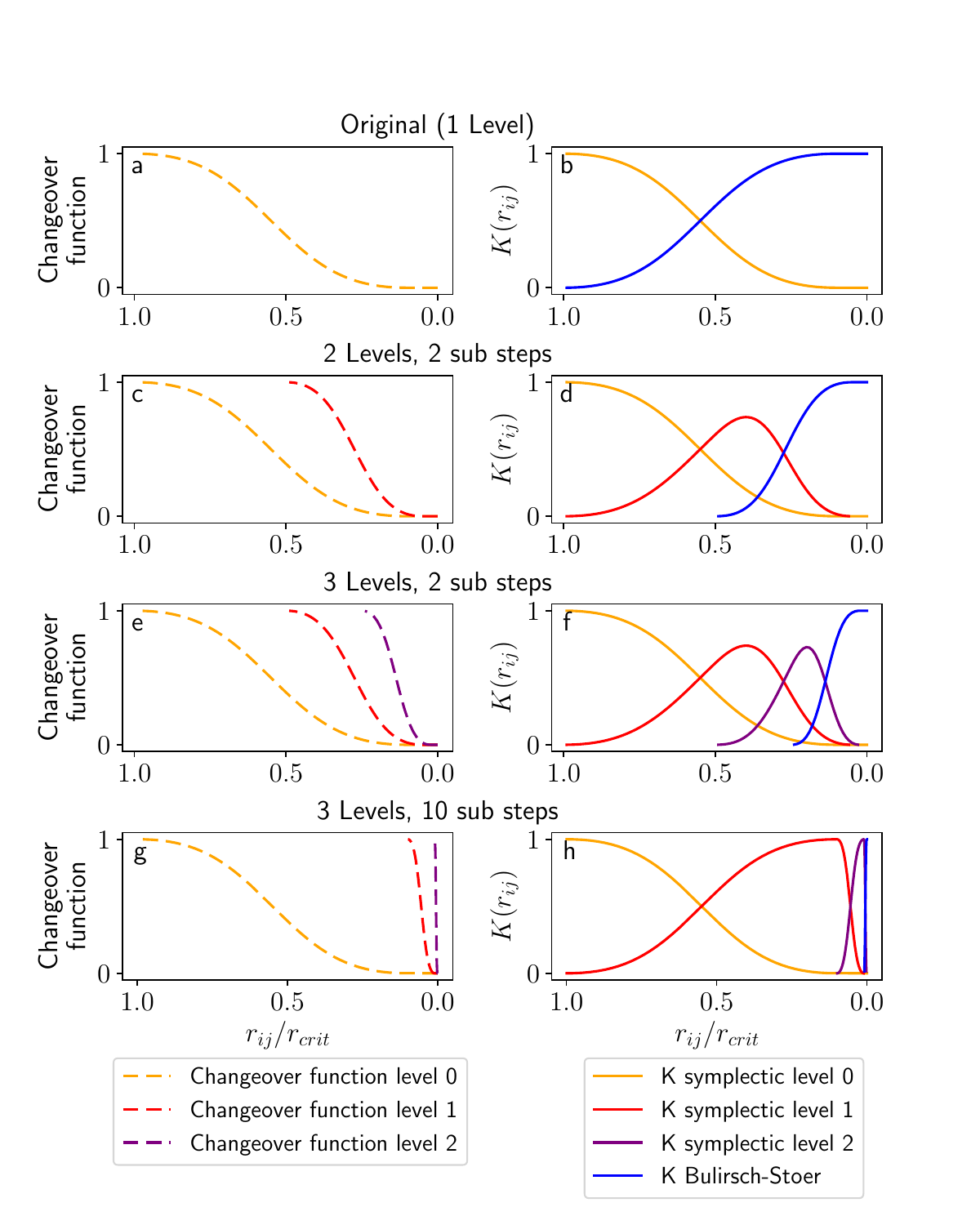}
\caption{Panels (a) and (b): original hybrid symplectic method, a changeover function switches smoothly from K symplectic to K Bulirsch-Stoer. \\
Panels (c) and (d): a second changeover function is included. The basic symplectic step is divided in the first level into two sub-steps. The second level is the Bulirsch-Stoer method. \\
Panels (e) and (f): two additional changeover functions are included for a three-level scheme with two sub-steps in each level.\\
Panels (g) and (h): two additional changeover functions are included for a three-level scheme with 10 sub-steps in each level.
}
\label{fig:K} 
\end{figure}

The original hybrid symplectic integrator described in Section \ref{sect:HybridSymplectic} consists of two levels: the basic level (level 0), which is the symplectic integrator with a fixed step size wherein bodies revolve around the central mass, and the Bulirsch-Stoer method (level 1), which is used for close-encounters between bodies other than the central mass. A changeover function smoothly switches between the two levels. This method can be extended by introducing a number of intermediate levels between the fully symplectic regime and the Bulirsch-Stoer regime: instead of switching directly to the Bulirsch-Stoer method another symplectic integration step with a reduced time step can be applied. This is desirable because the Bulirsch-Stoer steps are computationally expensive and should be used as a last resort. With the introduction of an additional level, a second changeover function can then smoothly switch between the new additional shorter symplectic step and the Bulirsch-Stoer method. In this way, the second argument in the critical radius (Equation \ref{rcrit}), $n_2\cdot v \cdot dt$, gets reduced in the higher levels because of the smaller time step. The Bulirsch-Stoer method is called much later in the close-encounter phase, and therefore used much less frequently. This method can reduce the close-encounter group sizes requiring the Bulirsch-Stoer integration saving valuable computation time. In practice there can be multiple such intermediate levels. This new formalism lies in between of the original hybrid symplectic integrator and the 'multiple time step symplectic algorithm' as implemented in the SyMBA code \citep{DuncanLevisonLee1998}. While the described method in this work also uses multiple symplectic time steps, unlike SyMBA, the deepest level is still calculated with a direct $N$-body method. The reduction factor on the time step of the intermediate levels can be set freely by the user. Common options are two, four, or 10, but any other number is also possible. Note that in SyMBA, a factor of three is used by default. 
Since the changeover function is acting in the range of $1.0 < r_{ij}/r_{crit} < 0.1$
\footnote{The factors 1.0 and 0.1 are defined by the form of the changeover function in \citet[Equation 10]{Chambers99}.} the choice of splitting a level into 10 sub-steps leads to a special case where no more than two levels are active at the same time. Figure \ref{fig:K} shows four examples of different numbers of symplectic levels and sub-steps per level. The panels (a) and (b) are identical to the original hybrid symplectic integrator, with a single changeover function $K(r_{ij})$. In this case we define

\begin{eqnarray}
K_{\text{symplectic}} = K(r_{ij}) \\ \nonumber
K_{\text{Bulirsch-Stoer}} = 1 - K(r_{ij})
\end{eqnarray}

The panels (c) and (d) show an example with two levels and each level is split into two sub-steps. In this case we define

\begin{eqnarray}
& K_{\text{symplectic 1}} = K(r_{ij}, dt) \\ \nonumber
& K_{\text{symplectic 2}} = (1 - K(r_{ij}, dt )) \cdot K(r_{ij}, dt / 2) \\ \nonumber
& K_{\text{Bulirsch-Stoer}} = (1 - K(r_{ij}, dt )) \cdot (1 - K(r_{ij}, dt / 2)),
\end{eqnarray}
with the base time step $dt$.

Panels (e) and (f) show three levels with two sub-steps, and the panels (g) and (h) show three levels with 10 sub-steps In all examples, the Bulirsch-Stoer method is only called at the blue lines.

An important detail is that we still use the same critical radius definition (equation \ref{rcrit}) as in the original algorithm. The only difference is that the time step is reduced in the higher levels, which leads to a smaller critical radius. In principle, the algorithm could also be used to reduce the Hill radius part of equation \ref{rcrit}, as the SyMBA method does \citep{DuncanLevisonLee1998}, but this is not done in the current version of GENGA. The reason is that, inside the Hill sphere of a body, the trajectory of a third object is no longer dominated by the central body, but is rather described by three-body dynamics.  Therefore, using a Kepler drift about the central body to advance the orbit is not very effective. A more detailed description of the algorithm is given in Appendix \ref{sect:algorithm}.

Figure \ref{fig:SLevelsTime} shows a performance comparison of different symplectic levels for an increasing number of planetesimals and for two different GPUs. The legend in the left panel indicates the number of levels and the number of sub-steps per level. The time is normalized to the original hybrid symplectic integrator (black dashed line). The initial conditions for the simulations are disks of $N$ planetesimals with a total mass of 5 $M_{\oplus}$ orbiting the Sun between 0.5 au and 4 au. The results show that using more symplectic levels can speed up the simulation, especially at high $N$. The results also show that there is not a unique best choice for all GPU types, and for every GPU an individual best choice must be found. However, both GPUs show that the symplectic levels only speed up the simulations when $N \gtrsim 4096$.

In practise, a good balance between the number of levels and the number of sub-steps must be found. Using a higher number of levels requires more memory to store all close-encounter pairs of each level, and using a higher number of sub-steps increases the amount of kernel calls needed and therefore can increase the total kernel overhead time. The fastest configuration will depend on the GPU device used. Therefore, we recommend testing different configurations before running long-term simulations in order to find the best option. In a future version of GENGA, this process could also be automated by a self-tuning routine at the beginning of the simulation.

\begin{figure}
\includegraphics[width=1.0\linewidth]{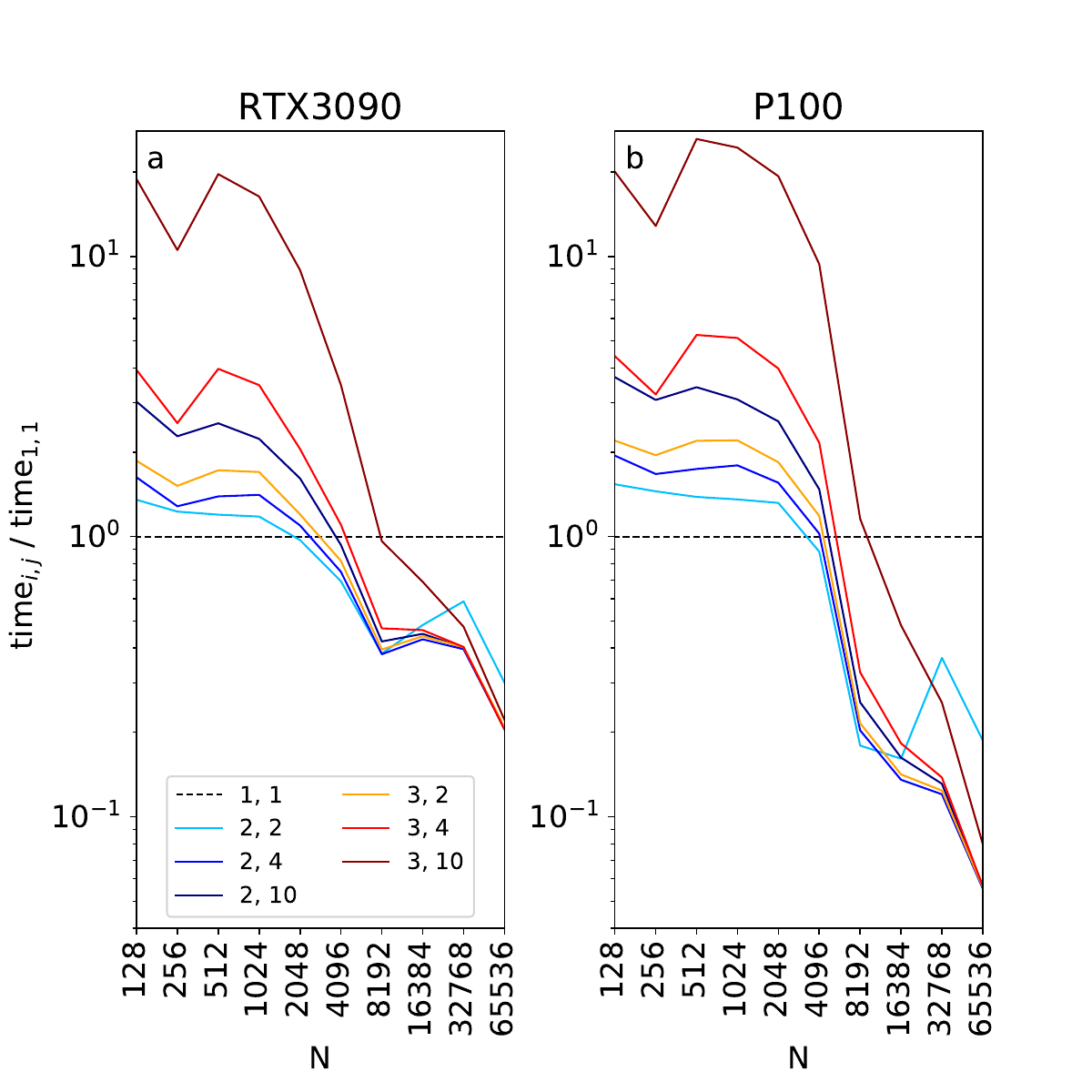}
\caption{Performance test of higher symplectic levels. The shown time is normalized to the time of the original hybrid symplectic integration method (1, 1). The legend indicates the number of symplectic levels, followed by the number of sub-steps per level. The initial condition is a disk of $N$ planetesimals, distributed between 0.5 au and 4 au, with a total mass of 5 $M_{\oplus}$. For a large number of bodies, using more symplectic levels can speed up the integration significantly. The comparison between different GPU types indicates that every GPU has a different optimal choice of parameters. For $N$ $<$ 4096 the choice of 1,1 is generally the fastest.
}
\label{fig:SLevelsTime} 
\end{figure}

\subsection{Example I: High-resolution terrestrial planet formation}
\label{sect:ExampleI}
\begin{figure*}[ht]
\includegraphics[width=1.0\linewidth]{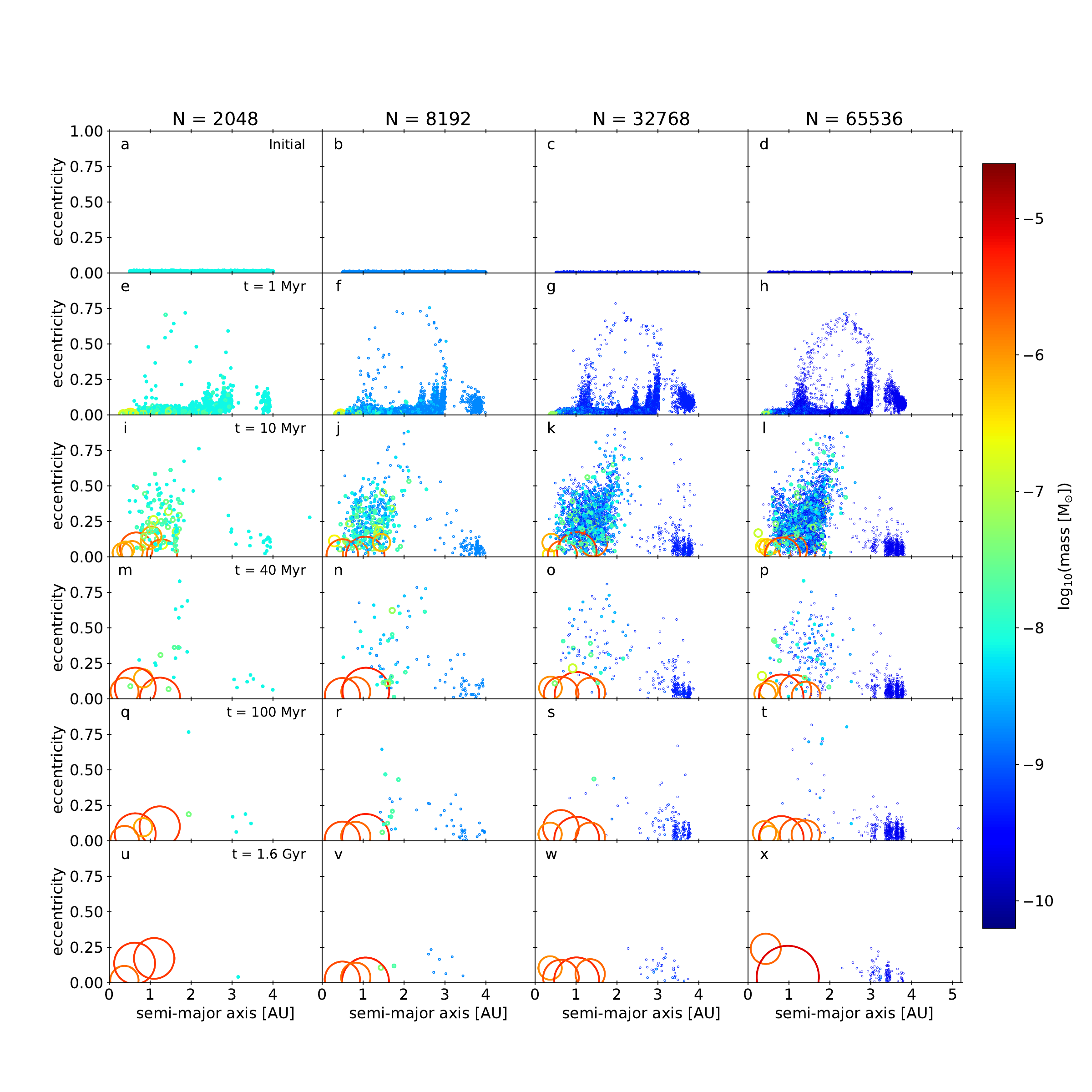}
\caption{Terrestrial planet formation starting with $N$ planetesimals with a total mass of 5 $M_\oplus$ and including a gas disk, Jupiter, and Saturn. Sweeping secular resonances from Jupiter and Saturn scatter planetesimals to the inner part of the disk, where terrestrial planets can form within 20-30 Myr. The final configuration of the formed planets is dependent on stochastic processes and can be different in each individual simulation. By including more and smaller planetesimals, the structure of an asteroid belt becomes visible. To achieve realistic masses in the asteroid belt, still much more and smaller particles must be used. The color and the size of the dots both represent the mass of the planetesimals. The gas giants are not shown in this Figure.}
\label{fig:aeiGas2} 
\end{figure*}

We use the described higher order changeover mechanism to
test the formation of the terrestrial planets via planetesimal collisions, with different particle resolutions. We use a disk of $N$ equal-mass planetesimals, distributed between 0.5 au and 4 au with a total mass of 5 $M_\oplus$, with the number of planetesimals varying between 2048 and 65,536. We include the gas giants Jupiter and Saturn on their current orbits and use a gas disk to simulate planetesimal migration. The sweeping secular resonances from the gas giants and the gas disk start to scatter the planetesimals between $\sim$2 au and $\sim$3.5 au to the inner part of the system, where terrestrial planets can form within 20-30 Myr. After $\sim$100 Myr, the terrestrial planets have accreated most of the surrounding planetesimals but, in some cases, they can still collide with each other. Snapshots of the simulations are shown in Figure \ref{fig:aeiGas2}. The mechanism of the sweeping secular resonances and more details of the used gas disk are given in \cite{Woo2021}.
The final configuration of the formed terrestrial planets are affected by the chaotic nature of the $N$-body problem and therefore depend on stochastic processes (e.g., in which order floating point numbers are processed (see Appendix \ref{sect:SerialGrouping})) so that the observed differences in the figure are not only caused by using different particle sizes. However one can see clearly that, by using an increased mass resolution, the structure of the asteroid belt becomes visible. In this test case, we include the gas giants already at the beginning of the simulation. Inserting them at a later time, or by using a growth track for their masses, would affect the location and structure of the remaining asteroid belt. Also, the initial distribution of the planetesimals strongly affects the final distribution of the asteroid belt. The execution time on a Tesla P100 GPU of this example is listed in Table \ref{tab:TimeExampleI}.

In order to find the best configuration for the number of levels and the number of sub-steps per level, we run each configuration for 1000 time steps before the simulation starts, and choose the fastest option. We repeat this procedure periodically as the number of particles gets reduced.

\begin{table}[ht]
\caption{Execution time of example I in months on a Tesla P100 GPU$^a$}
\label{tab:TimeExampleI}
\begin{tabular}{c | c | c | c | c}
 $N$ & $<10^6$yr & $10^6 - 10^7$yr & $10^7 - 10^8$yr & $10^8 - 10^9$yr \\
 \hline
2048 & 0.02 & 0.14 & 0.36 & 1.47 \\
4096 & 0.03 & 0.20 & 0.38 & 2.22 \\
8192 & 0.10 & 0.31 & 0.48 & 2.14 \\
16,384 & 0.65 & 0.68 & 0.57 & 2.86 \\
32,768 & 0.90 & 1.22 & 0.88 & 3.07 \\
65,536 & 2.76 & 2.60 & 1.30 & 3.78 \\
\end{tabular}
\footnotesize{$^a$ The execution time shown is the total run time of the simulations. The computer cluster used had a wall time limit of 24 hr for each job. When the wall time limit was reached, GENGA automatically saved the vectors at the final time step before it is terminated. Therefore, continuing a simulation on such a cluster takes only $\sim $5-30 s per restart. Restarting simulations periodically can improve the performance slightly because the self-tuning step is executed more often (see section \ref{sect:self-tuning}) and each segment of the simulation will run faster as a result.}
\end{table}

\subsection{Example II : High-resolution planet formation without gas giants}
\label{sect:ExampleII}
\begin{figure*}
\includegraphics[width=1.0\linewidth]{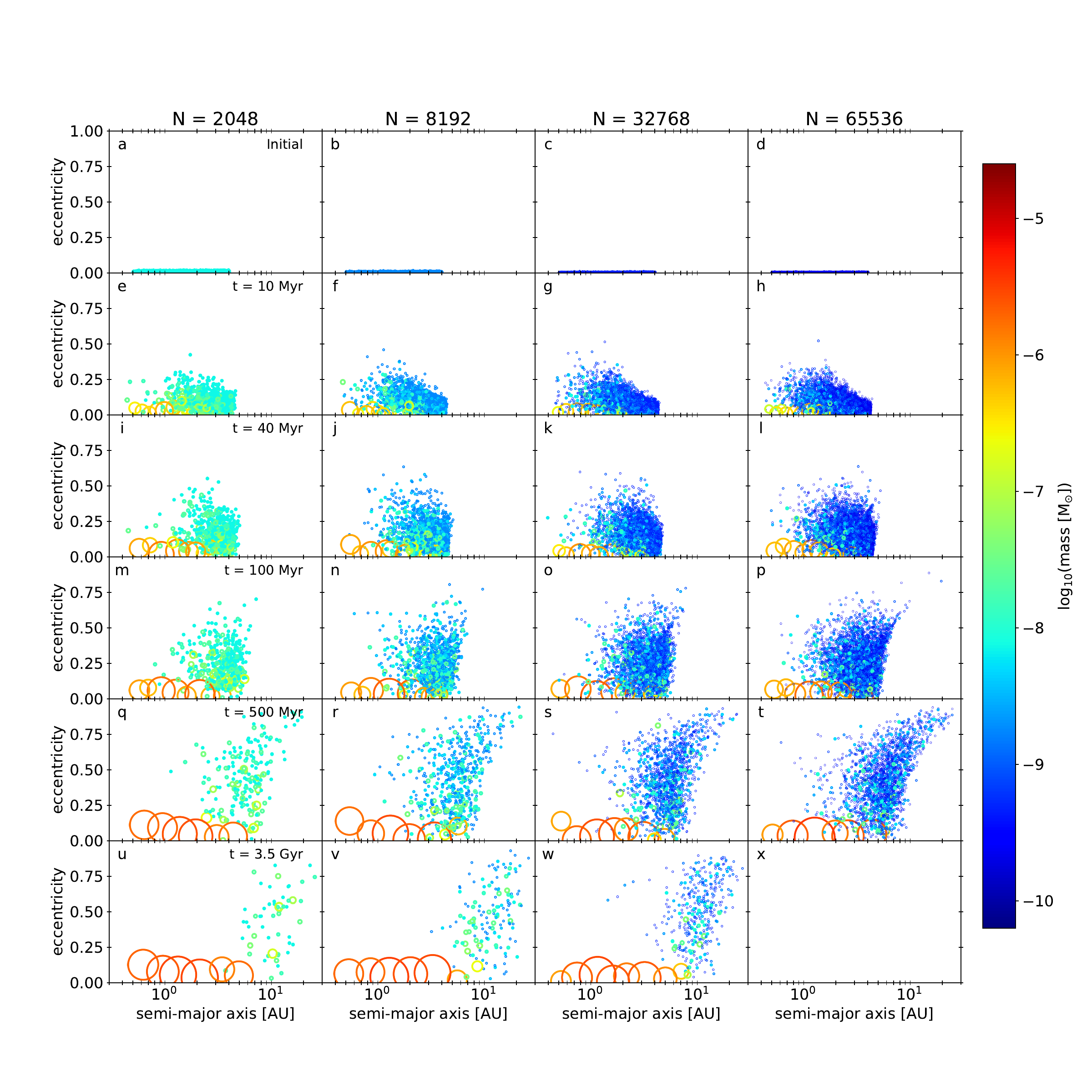}
\caption{Planet formation starting with $N$ planetesimals with a total mass of 5 $M_\oplus$
and without any gas giants, which significantly slows the planet formation process \citep{QuintanaLissauer2014}. Planetary embryos are first formed at the inner disk, where they scatter the planetesimals to a larger semi-major axis and more eccentric orbits. At the outer edge, planet formation still continues after more than 3 Gyr. The snapshots at 100 Myr (panels (m) to (p)) and 500 Myr (panels (q) to (t)) show that smaller planetesimals can be scattered away earlier than more massive planetesimals. The snapshot at 3.5 Gyr (panels (u) to (w)) indicate that, for smaller planetesimals, the planet formation process at the outer edge of the disk is not yet finished after 3.5 Gyr. Panel (x) is empty because this simulation reached only 500 Myr. The color and the size of the dots both represent the mass of the planetesimals.}
\label{fig:aei2} 
\end{figure*}

We repeat the planet formation simulations of the previous section without any gas giants and without a gas disk. Not including the gas giants in the system has a huge impact on the planetary formation process. The planetesimals are not exposed to large mean motion or secular resonances and therefore their eccentricities remain very low for a long time. This also means that the influence of a gas disk is less pronounced and has only little effect on the final configuration. This is the expected result because the formation process takes longer than the lifetime of the gas disk; we confirm this in test simulations. Therefore, we do not include the gas disk in this section and the simulation is only using pure Newtonian gravity. In Figure \ref{fig:aei2} we display the result of four simulations with a different number of planetesimals. The total amount of mass is always 5 $M_\oplus$. Without the gas giants, the planet formation process is much slower and growth is inside out. The formed planetary embryos or planets scatter the surrounding planetesimals to a larger semi-major axis and a more eccentric orbit. With this mechanism, planets can also be formed outside of the original planetesimal disk. The figure shows that smaller planetesimals are scattered outside faster than larger planetesimals, and that planet formation continues over a longer time. After 3.5 Gyr, planet formation is still not finished in the outer part of the disk. The execution time on an Tesla P100 GPU of this example is listed in Table \ref{tab:TimeExampleII}. The execution time is larger than in the example I, because more planetesimals remain in the simulation.

\begin{table}[ht]
\caption{Execution time of example II in months on a Tesla P100 GPU}
\label{tab:TimeExampleII}
\begin{tabular}{c | c | c | c | c}
 $N$ & $<10^6$yr & $10^6 - 10^7$yr & $10^7 - 10^8$yr & $10^8 - 10^9$yr \\
 \hline
2048 & 0.02 & 0.12 & 0.55 & 2.14 \\
4096 & 0.04 & 0.23 & 0.87 & 2.32 \\
8192 & 0.09 & 0.45 & 1.54 & 3.57 \\
16,384 & 0.22 & 0.91 & 2.68 & 5.04 \\
32,768 & 0.68 & 2.32 & 4.93 & 7.82 \\
65,536 & 2.41 & 7.22 & 9.51 & -
\end{tabular}
\end{table}

\section{Improved Collision and Encounter options}
\label{sect:Coll+Enc}
In section \ref{sect:closeEncounters} we describe an improved method to handle a large number of close-encounters numerically in an $N$-body integrator. We also describe updates to the collision handling between particles and new diagnostic tools for keeping track of physical close-encounters.  
\subsection{Collision handling}
\label{sect:Collisions}
A collision between two bodies happens when their physical radii overlap ($r_{ij} < R_i + R_j$), and the two bodies are merged into a single body. The current version of GENGA treats collisions as perfectly inelastic mergers by conserving linear momentum. Physically, a part of the potential and kinetic energy is transformed into internal energy that will be dissipated as heat. GENGA keeps track of this internal energy, such that the overall energy of the system is conserved. Angular momentum is conserved by a transformation into the spin of the new body. The position and velocity of the new body are given by

\begin{equation}
   \mathbf{x}_{\rm new} = \frac{\mathbf{x}_i m_i + \mathbf{x}_j m_j}{m_i + m_j}
\end{equation}
and

\begin{equation}
   \mathbf{v}_{\rm new} = \frac{\mathbf{v}_i m_i + \mathbf{v}_j m_j}{m_i + m_j},
\end{equation}
where $i$ and $j$ indicate the indices of the two colliding bodies.
The spin $\mathbf{S}$ of the new body is calculated as

\begin{equation}
   \mathbf{S}_{\rm new} = \mathbf{S}_i + \mathbf{S}_j + \mathbf{L}_{\rm ij},
\end{equation}
with
\begin{equation}
\mathbf{L}_{ij} = \frac{m_i m_j}{m_i + m_j} \left( \mathbf{r}_{ij} \times \mathbf{v}_{ij} \right)
\end{equation}
being the orbital angular momentum. The new radius $R$ is set by conserving the mass and by mixing the densities of the two bodies

\begin{equation}
   R_{\rm new} = \left( R_i^3 + R_j^3 \right)^{1/3}
\end{equation}
and the new mass is just the sum of the masses
\begin{equation}
   m_{\rm new} = m_i + m_j.
\end{equation}

In order to keep track of the energy conservation, we add the lost kinetic and potential energy from collisions, ejections, and that caused by gas drag into a quantity $U$. This quantity includes the energy loss from all particles together and not just from single particles. The lost energy $U$ is not directly related to a physical quantity, however it contains the change of the inner energy of all particles and the spin energy caused at collision. 

When two particles $i$ and $j$ collide, then $U$ is increased by
\begin{equation}
   \Delta U = \frac{1}{2} \frac{m_i m_j}{m_i + m_j} v_{ij}^2 - G \frac{m_i m_j}{r_{ij}}.
\end{equation}

The index of the new merged body is set according to
\begin{itemize}
 \item either the index of the more massive body or
 \item if both bodies have an equal mass, then the lower index of the bodies $i$ and $j$.
\end{itemize}

\subsection{Collision precision}
The collision process is resolved during the Bulirsch-Stoer direct integration of the bodies in a close-encounter with discrete time steps. Therefore, a collision is generally not detected at the exact collision time, but rather when the two particles already overlap by a small amount. In some situations, it can be necessary to compute the collision time more precisely and to extract the coordinates of the two bodies at the exact collision time. GENGA offers the option to control the collision precision by a user parameter called 'Collision Precision.' This parameter sets the tolerance of the collision detection and is set in units of a radius fraction, i.e.
\begin{equation}
\frac{(R_i + R_j) - r_{ij}}{R_i + R_j} < \text{tolerance}.
\end{equation}
It is also possible to define if collisions should be reported with slightly overlapping positions, or if positions should be reported shortly before the exact collision time. The latter option is needed when a bouncing collision model is implemented.

An example of a collision is shown in Figure \ref{fig:Collision}. The orange circles indicate the positions and sizes of the bodies at the regular time steps. During a close-encounter phase, the time steps are refined within the Bulirsch-Stoer method, shown in thin green color. When a low precision is used, then the collision is reported at the time step when the two bodies start to overlap. When a high collision precision is used, then the contact point is refined until the tolerance is reached.
 
\subsection{Trace back collisions}
\begin{figure}
\includegraphics[width=1.0\linewidth]{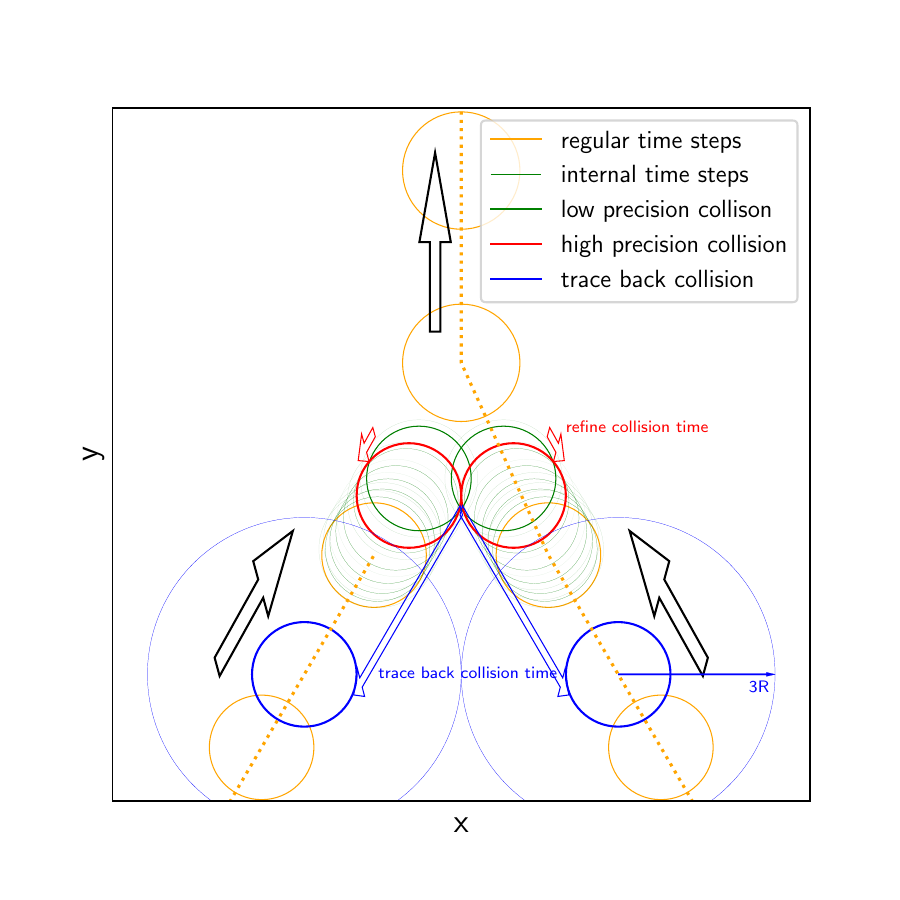}
\caption{Example of two colliding bodies. The orange circles show the regular time steps of the integration, the thin green circles the internal reduced time steps during the Bulirsch-Stoer close-encounter method. The precision of the resolved collision time can be set by a user parameter. The green color indicates a low precision, the red color indicates a high precision. Collisions can also be traced back to a time before the actual collision happens, where the two bodies are still separated by an increased radius. In blue color is shown such a back-traced collision at a position where the distance between the two bodies corresponds to three times the sum of their radii.
}
\label{fig:Collision} 
\end{figure}

For certain applications, it is necessary to extract the coordinates of two colliding bodies at an earlier time, when the two particles are still separated by two or three times their radii. When this is the case, collisions are resolved with an external method, e.g. a model that resolves the inner structure of the bodies. GENGA has the option to back trace collisions to a time before the actual collision happens. In order to do that, we save the current time step where the collision is detected and apply a backward time step. During this backward time step, we increase the radii of the two affected bodies by the desired fraction, and we report the new detected collision positions. After that, GENGA can continue at the saved time step as usual. When multiple collisions happen at the same time step, then we have to resolve each of them independently, because the increased radii of two bodies can affect the dynamics of other particles as well. An example of a back-traced collision is shown in Figure \ref{fig:Collision} in blue color.

\subsection{Stop at collision time}
When an external code is used to resolve collisions in more detail (e.g., \cite{Timpe2020}), it is necessary to stop the entire simulation at the time of the first detected collision, or at the earliest back-traced collision time. It is important that all other particles, which are not involved in the collision or in a close-encounter, are also integrated backward in time self consistently. GENGA supports this option to stop a simulation and creates a separated collision output file containing all bodies at the time of the first collision. This file can then be used to resolve the collision externally and to create a new initial condition file to continue the $N$-body integration with GENGA until the next collision is detected. 

\subsection{Report Close Encounters}
GENGA supports the option to store the coordinates of all detected close-encounters in a file. In this context, a close-encounter is not defined as the usual numerical close-encounter (Equation \ref{rcrit}), but rather when the separation $r_{ij}$ between the bodies $i$ and $j$ is smaller than the sum of the increased physical radii:
\[
r_{ij} < f (R_i +  R_j),
\]
where $R_i$ and $R_j$ are the physical radii of the two bodies, and the factor $f$ a user parameter. An encounter event is reported at the moment of closest approach, which typically happens only once per orbit. In order to find all such close-encounters, the critical radius $r_{\rm crit}$ (Equation \ref{rcrit}) is automatically increased if necessary.  An example of an application where this option can be used is shown in section \ref{sect:Hebe}.

\subsection{Stop at Encounters}
It is possible to stop a simulation automatically when an encounter between two bodies occurs. In this context, a close-encounter is defined when the separation $r_{ij}$ between the bodies $i$ and $j$ is smaller than the sum of the increased Hill radii:
\[
r_{ij} < g (R_{H,i} + R_{H,j}),
\]
where $R_{H,i}$ and $R_{H,j}$ are the Hill radii of the two bodies, and the factor $g$ is a user parameter.

\section{New tools and options in GENGA}
\label{sect:NewTools}
In the following section, we describe additional improvements and updates of GENGA since \cite{GrimmStadel2014}. These updates are 1) a new semi-active test particle mode, 2) a function to use predefined masses, 3) radii or orbital elements tables, 4) a self-tuning performance optimization procedure and 5) the support for AMD GPUs. More tools and functions are listed in the Appendix \ref{sect:AppB}, which includes a real-time visualization tool using openGL, a method to create exact reproducible results, an option for creating $a-e$ and $a-i$ grids of a simulation, and an option for a GPU buffer for outputs.   


\subsection{New semi-active test particles mode}
\label{sect:TP2}
We add a test particle mode for semi-active bodies. In this mode, test particles can interact with large bodies, but they do not interact with other test particles. This mode can be used when simulating small particles whose total gravitational potential influences the orbits of larger planets as used in \cite{Quarles2019}, for example. Since test particles in the semi-active mode can also influence large particles, the close-encounter groups must be calculated the same way as in the fully interactive mode. When simulating planet formation by merging small planetesimals, then the test particle mode should not be used, because it would prevent the collisions of planetesimals and therefore the formation of larger bodies through the merger of smaller ones. An exception could be made when the collisions of small particles are treated statistically. Test particles do not have to be massless, the user can specify a mass threshold for treating particles as test particles or as fully interactive particles.

\begin{figure}
\includegraphics[width=1.0\linewidth]{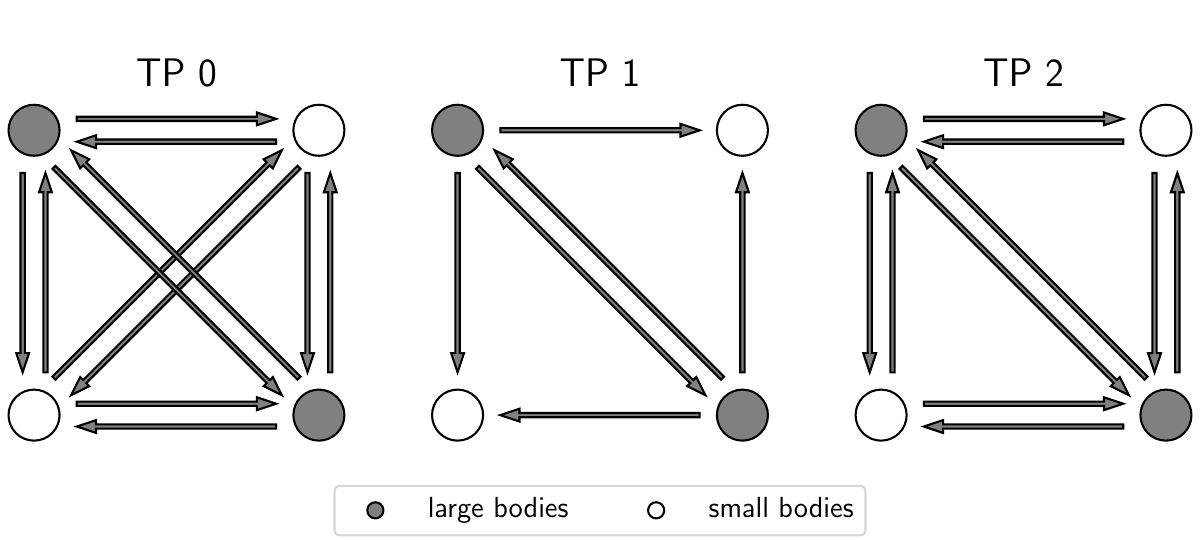}
\caption{Schematic of the different test particle modes (TP). The left panel shows the full interactive mode, where each particle interacts with all the other particles. The middle panel shows the test particle mode 1, where only large particles interact with other particles. The right panel shows the test particle mode 2 for semi-active particles, where small particles do not interact with other small particles but with other large particles. The arrow in the Figure indicates all gravitational force terms that are calculated. 
}
\label{fig:TestParticles} 
\end{figure}

The different test particles modes of GENGA are illustrated in Figure \ref{fig:TestParticles} and can be summarized as follows:
\begin{itemize}
    \item Fully interactive mode (TP 0).\\
    All bodies interact with all the other bodies. This is the default.
    \item Test particles mode (TP 1). \\
    Test particles do not interact gravitationally with other particles, except when they collide with large bodies, then their momentum is added to the final momentum of the new body.
    \item Semi-active test particle mode (TP 2). \\
    Test particles do not interact with other test particles. They interact with large bodies the same way as in the fully interactive mode. 
\end{itemize}

The performance of the test particles modes is described in more detail in section \ref{sect:TestParticlesPerformance}.

\subsection{Usage of predefined coordinates, mass or, radius tables}
\label{sect:SetElements}
The set-elements option in GENGA permits users to modify the orbital parameters, mass, or radius of bodies, according to a predefined data file. This can, for example, be used to fix the position of a planet to a certain location and make chaotic evolution more or less reproduceable, or to use a pre-calculated mass-radius evolution table of a planet and let the planet grow with time. When Keplerian elements are provided, then the code converts the Cartesian coordinates during the simulation into Keplerian elements, modifies them according to the table, and converts them back into Cartesian coordinates. At each time step, GENGA interpolates temporally between given values in the table by using a cubic interpolation scheme. An application using a mass and radius table is presented in Section \ref{sect:Jupiter}.

\subsubsection{Example: planetesimal accretion and satellite capture by Jupiter}
\label{sect:Jupiter}
\begin{figure}[h]  
  \begin{minipage}[b]{1.0\linewidth}
    \includegraphics[width=0.99\linewidth]{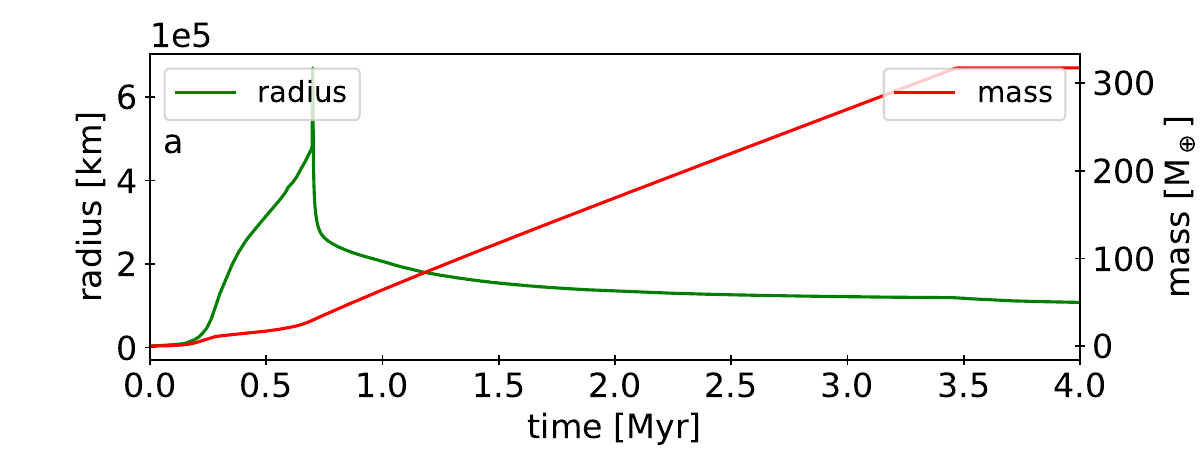} 
    \includegraphics[width=0.99\linewidth]{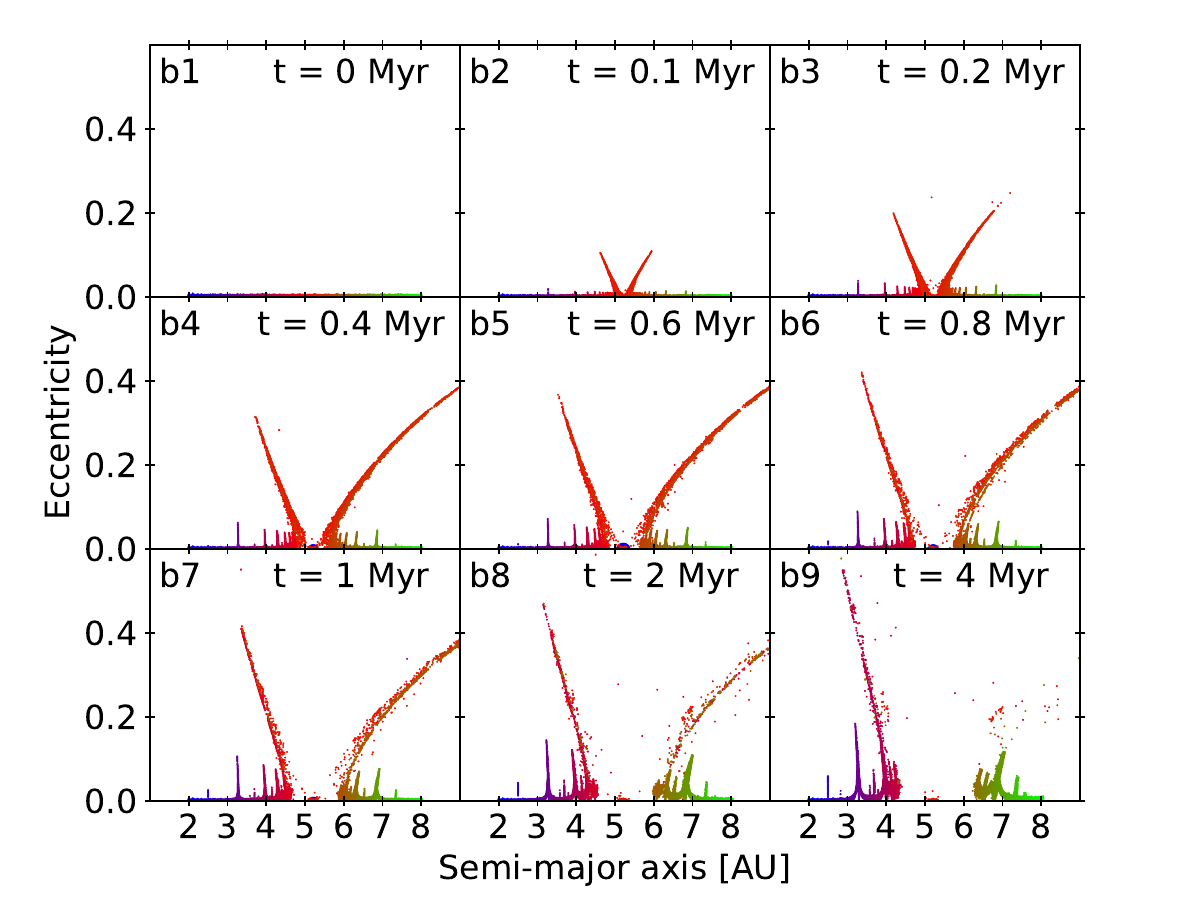} 
    \includegraphics[width=0.99\linewidth]{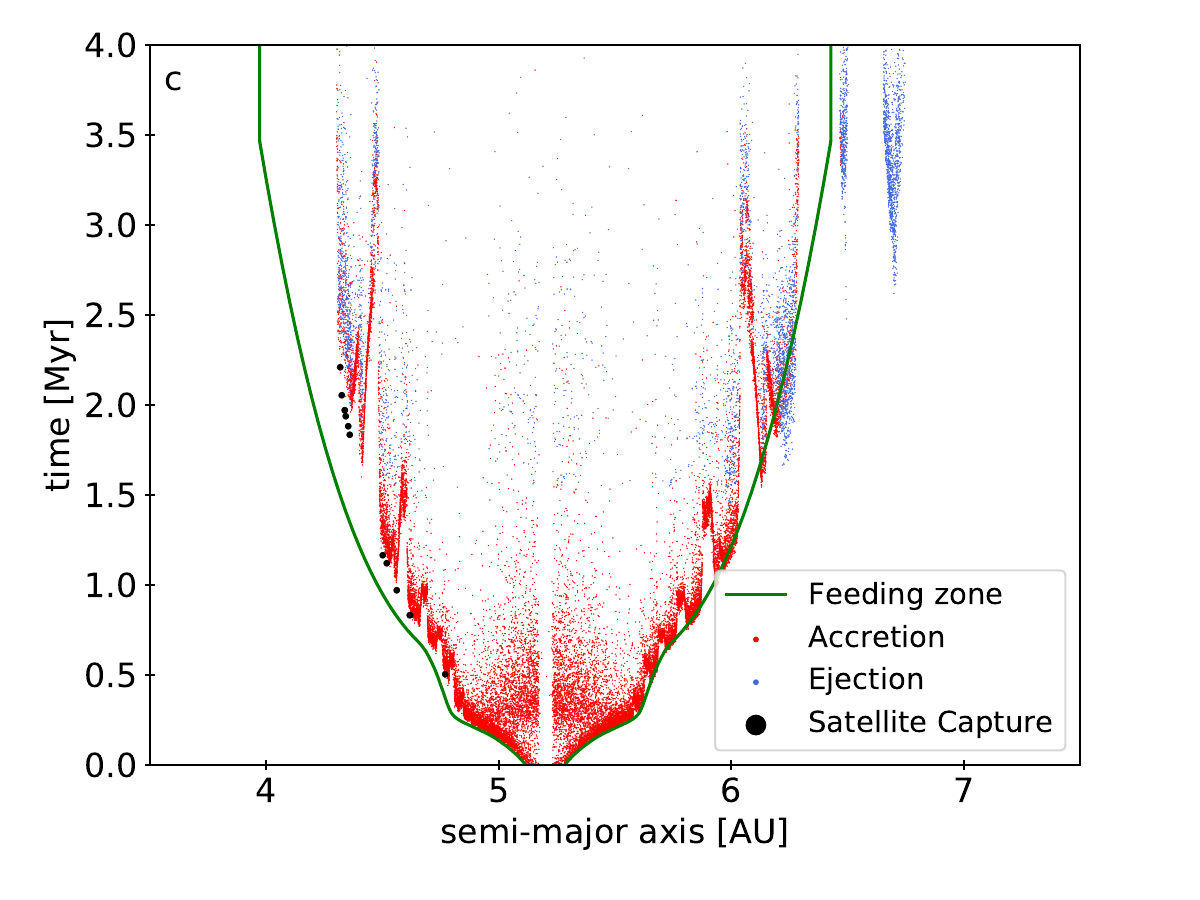} 
  \end{minipage}
  \caption{Example of a growing Jupiter simulation by using a predefined mass and radius table. The top panel shows the time evolution of the mass and planetesimal capture radius of a Jupiter-like planet. The middle panel shows the evolution of a planetesimal disk, distributed between 2 and 8 au, where the color indicates the original distance to the Sun. One can see how the gas giant removes all particles from the mean-motion resonance locations. The bottom panel shows the theoretical feeding zone of the giant planet in green color and all accretion and ejection events of the simulation. The x-axis indicates the original semimajor axis of the planetesimals at the beginning of the simulation. The black dots indicate satellite capture events, where some planetesimals evolve from a circumsolar orbit into a circumplanetary orbit.}
  \label{fig:JupiterTracks}
\end{figure}

In this example, we simulate the planetesimal accretion of a forming Jupiter-like planet \citep{Pollack1996,Alibert2005}. We use a pre-calculated table with the planet mass and planetesimal capture radius of the growing gas giant, depicted in the top panel of Figure \ref{fig:JupiterTracks}. This data was obtained in a similar way as described in \citet{Mordasini2012b}, i.e. with a classical 1D giant planet formation model based on the core accretion paradigm akin to \citet{Pollack1996}. The planetesimal capture radius is calculated as in \cite{InabaIkoma2003}. This means that this model combines the solution of the planetary interior structure equations to get the gas accretion rate with the accretion of planetesimals based on a rate equation to calculate the core accretion rate \citep{Pollack1996, Alibert2005}.
The growing gas giant is fixed at a semimajor axis of 5.2 au and reaches its final mass and radius after about 3.5 Myr. We embed the gas giant into a disk of 100,000 planetesimals, distributed uniformly between 2 and 8 au. The planetesimals are treated as massless test particles. At the beginning, the gas giant accretes nearby planetesimals and it starts to excite the eccentricities of planetesimals located in mean-motion resonances and those that get within a few Hill radii. When the excited planetesimals reach a distance of 1000 au to the Sun, they are removed from the simulation and reported as ejected particles. The middle panel of Figure \ref{fig:JupiterTracks} shows the time evolution of the semi-major axis and eccentricities of the planetesimals. The color in the middle panel indicates the original semi-major axis of the particles. After the gas giant has reached its final mass, it has cleared the planetesimal disk between 4.5 and 6.5 au, which roughly corresponds to the theoretical feeding zone. Since the mass of the gas giant increases with time, it is possible that some planetesimals are captured as satellites. In our application, we have observed 11 captured satellites from the inner edge of the feeding zone. Satellite captures from the outer edge of the feeding zone can also happen, but are more rare. The bottom panel of figure \ref{fig:JupiterTracks} shows all observed satellite capture events together with all planetesimal accretion and ejection events. At the beginning of the simulation, when the mass of the gas giant is still small, it collects all surrounding planetesimals within the feeding zone. But while the planet grows in mass, it is not able to accrete more objects, but instead the planet starts to scatter them away until they leave the solar system or collide with other planets \citep{Levison2010}. 

The simulation was run for four million years.
The final number of accreted, ejected, and remaining planetesimals is 28,483 (28\%), 4915 (5\%), and 66,602 (67\%), out of the initially 100,000 (100\%). This simulation does not include the effect of gas drag on the planetesimals. However, tests indicate that the gas drag can be very important, and that the structure of the gas disk around Jupiter can change the number of accreted and ejected significantly. 
In this paper, this simulation mainly serves as an illustration of the possibility to use predefined mass and radius tables in GENGA. We note, however, that there is an ongoing discussion in the literature about the efficiency of planetesimal accretion by forming giant planets (see \citealt{Podolak2020} and references therein).   

It is also interesting to note that our observed satellite capture events could by applied to exoplanetary systems as well. For instance, two potential exomoon candidates are discussed (Kepler 1625b-i \citep{TeacheyKipping2018} and Kepler 1708b-i \citep{Kipping+2022}), which could have formed in a similar scenario \citep{HamersPertegiesZwart2018}.

\subsection{Use self-tuning kernel parameters}
\label{sect:self-tuning}
All GPU kernels need to be configured with a specific number of threads per thread block and a specific number of thread blocks, and the performance of the kernel can depend on this choice. Furthermore the best choice of kernel parameters can also depend on the initial conditions and the specific GPU. Therefore it is impossible to know beforehand which choice of kernel parameters leads to the best performance. To find the best choice ensuring the simulation runs as fast as possible, we implement a self-tuning routine in GENGA at the beginning of the integration. This routine tests different configurations and then adopts the best kernel parameters. The self-tuning routine can be disabled by a user parameter if needed but it is not recommended. By not using the self-tuning and using non-optimal kernel parameters, the performance penalty can be less than $1\%$ in some cases or more than $100\%$ in other cases, depending on the GPU type and the used initial conditions.

\subsection{Support for AMD GPUs}
In the last few years, AMD has developed an equivalent programming language to CUDA called HIP. Today, HIP supports most of the functionalities of CUDA, but not everything. An important difference to CUDA is that the warp size is not fixed to 32: it can be either 32 or 64. We updated GENGA such that it can run with different warp sizes, and we developed a translation tool to port GENGA from CUDA to HIP. We have successfully tested the HIP GENGA version on an AMD Radeon VII and an AMD Instinct MI100 GPU. While the results of simulations agree between the NVIDIA and the AMD GPUs, the performance of small simulations on AMD cards is not yet satisfying. The AMD GPUs seem to suffer much more from long kernel overhead time. This can probably be reduced by rearranging and combining different kernels into fewer kernel calls, but that is subject to future work.

\section{Performance}
\label{sect:Performance}
We measure the performance of the code by integrating a planetesimal disk with $N$ bodies, distributed between 0.5 and 4 au and a total mass of 5 $M_\oplus$. The number of close-encounters increases with the number of bodies. For less than 256 bodies, no close-encounters occur in this test. We use the best choice of the number of symplectic levels and sub-steps for each GPU type (see section \ref{sect:SLevels}). We use an NVIDIA GTX 980, NVIDIA GTX 1080, NVIDIA RTX 3090, and an AMD Radeon VII GPU, all installed in desktop machines. We further test the NVIDIA Tesla K20, NVIDIA Tesla P100, NVIDIA Tesla A100, and an AMD Instinct MI100 installed on HPC server systems in Switzerland, Norway, Finland, and Slovenia. Finally, we also test an NVIDIA Quadro T2000 GPU from a notebook machine. While the Tesla, Instinct, and the Radeon VII cards support fast double-precision computing, typically half the speed of single precision, the other GPUs only have an artificially reduced double-precision performance, typically only 1.5\%-4\% (1/64 to 1/24) of the single precision performance. 

\begin{figure}
\includegraphics[width=1.0\linewidth]{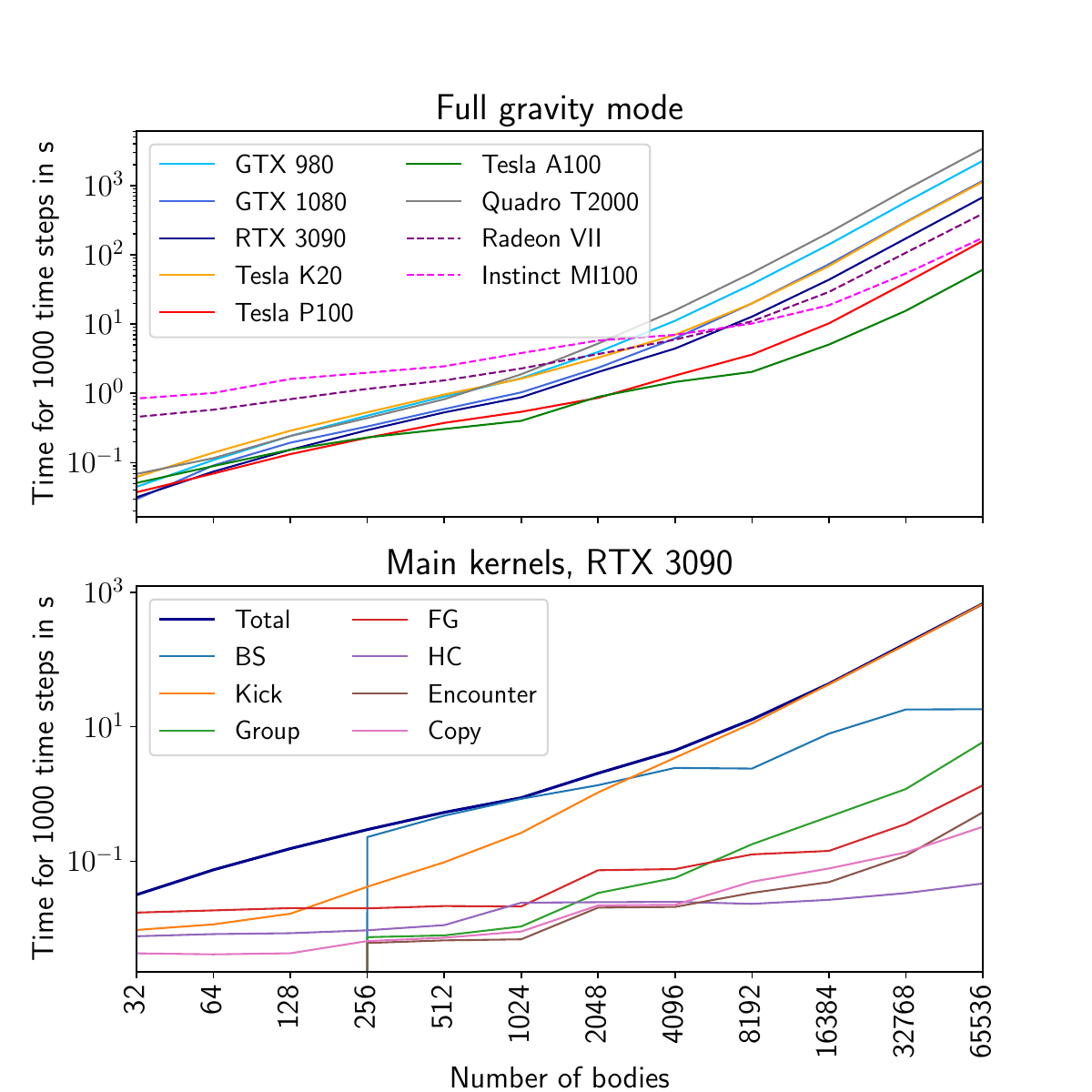}
\caption{Performance of the full gravity mode, for a set of $N$ massive bodies distributed in a disk between 0.5 and 4 au and a total mass of 5 $M_{\oplus}$. The GTX, RTX, and Radeon cards are installed in desktop machines, the Tesla and Instinct cards in a computer cluster and the Quadro card is integrated in a notebook machine. The GTX, RTX, Tesla, and Quadro cards are NVIDIA GPUS, the Radeon and Instinct cards are AMD GPUs.
The bottom panel shows the performance of the main kernels: BS (Bulirsch-Stoer routine for close-encounters), Kick (gravity calculation), group (close-encounter pairs grouping), FG (Kepler orbit solver), HC (Sun kick), encounter (close-encounter check) and copy (data transfer for close-encounter information). The difference between the shown kernels and the total time is caused mostly by the kernel overhead time.
}
\label{fig:Time} 
\end{figure}

The top panel of Figure \ref{fig:Time} shows the time for 1000 time steps as a function of the number of bodies for different GPUs. Since the Tesla GPUs support fast double-precision calculations, they are significantly faster for high-$N$ simulations. For low-$N$, the RTX 3090 card can benefit from more CUDA cores and a high clock speed of close to 2 GHz. The AMD GPUs show a good performance for large simulations, but they suffer from much more kernel overhead time than the NVIDIA GPUs for small simulations.
The bottom panel of Figure \ref{fig:Time} shows the performance of the most important kernels, measured on the RTX 3090 GPU. For low N, the FG (Kepler drift) kernel dominates the execution time, and also the total kernel overhead time and data transfer is important. For an intermediate number of bodies (256 $<$ $N$ $<$ 4096), the Bulirsch-Stoer integration dominates the full integration, while for high $N$ ($>$ 4096), the kick operation takes up most of the time. This result demonstrates clearly how the code can use different kernel optimizations, depending on the number of bodies. Note that the performance of the code also depends on the initial conditions, and especially how many close-encounters occur.

\subsection{Test particles}
\label{sect:TestParticlesPerformance}
To measure the performance of the test particle mode, we set up a particle disk between 0.5 and 4 au with zero inclination. The disk contains 16 massive particles which perturb the test particles as well as each other. This setup also leads to close-encounters between test particles and the massive bodies. We test the performance of the test particle modes 1 and 2 on the same GPUs as the full self-gravity tests.

\begin{figure}
\includegraphics[width=1.0\linewidth]{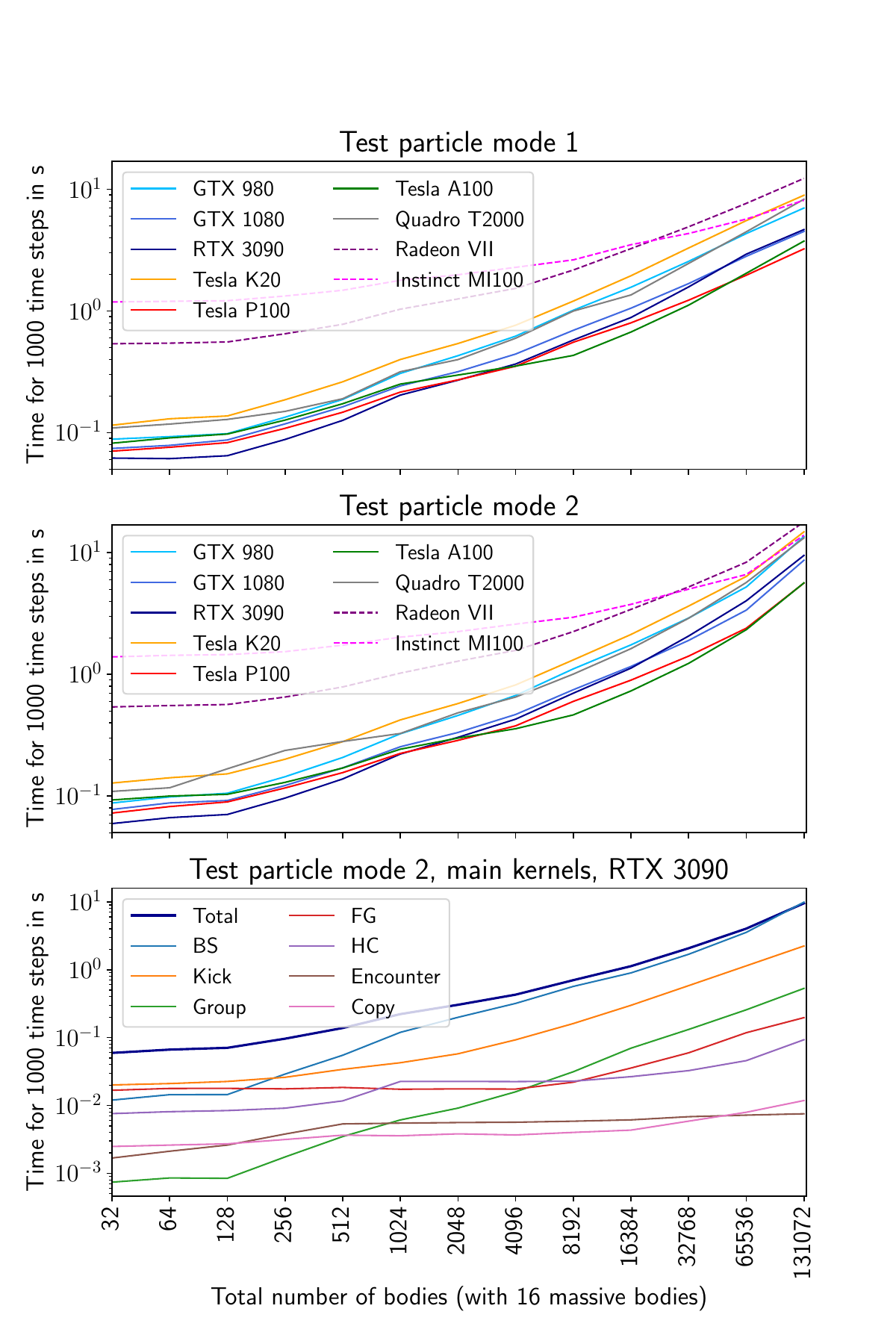}
\caption{Performance of the test particle modes 1 and 2, for a setup with 16 massive bodies and ($N$-16) test particles, embedded in a disk between 0.5 and 4 au.  The GTX, RTX and Radeon cards are installed in desktop machines, the Tesla and Instinct cards in a computer cluster and the Quadro card is integrated in a notebook machine. The GTX, RTX, Tesla, and Quadro cards are NVIDIA GPUS, the Radeon and Instinct cards are AMD GPUs. The bottom panel shows the performance of the main kernels: BS (Bulirsch-Stoer routine for close-encounters), Kick (gravity calculation), group (close-encounter pairs grouping), FG (Kepler orbit solver), HC (Sun kick), encounter (close-encounter check) and copy (data transfer for close-encounter information). The difference between the shown kernels and the total time is caused mostly by the kernel overhead time.
}
\label{fig:TimeTP} 
\end{figure}

In figure \ref{fig:TimeTP} we show the measured performance as a function of the particle number. For a high number of particles ($N$ $>$ 16,384), the Tesla P100 and Tesla A100 are clearly the fastest option, while for a low number of particles ($N$ $<$ 1024), the RTX 3090 is faster in this setup. The AMD GPUs show a good performance for high-$N$ simulations, but they suffer from much more kernel overhead time than the NVIDIA GPUs for low-$N$ simulations. In the bottom panel of Figure \ref{fig:TimeTP} we once again plot the performance of the main kernels. For a high number of particles, the Burlish-Stoer routine from the close-encounter phase clearly dominates the run time, followed by the kick operation. For a small number of particles, the kick operation and the FG routine for the Keplerian orbit calculation are comparable in run time. Also the total kernel overhead time is important for a low number of particles.

\section{Discussion and Conclusions}
We updated the GPU $N$-body code GENGA from its first invocation \citep{GrimmStadel2014} to be able to integrate high-resolution planet formation simulations by introducing a hierarchical close-encounter method with multiple changeover functions. We improved the performance of individual kernels by optimizing the parallelization and memory usage, and by adding a kernel self-tuning mechanism that finds the best kernel parameters for different GPU types. We added several non-Newtonian forces and a model for small-body collisions for meteorite dynamics. We included the option of using semi-active test particles, different options of setting the collision precision handling and added other tools like a real-time visualization with openGL.

With the described updates and optimizations, the performance of the new GENGA version can be up to two orders of magnitudes faster than the original methods reported in \cite{GrimmStadel2014}, depending on the initial conditions used. In the example simulations, we presented a planet formation simulation with a gas disk that ran longer than 1.5 billion years. And we presented a planet formation simulation without gas giants performed over 3.5 billion years that shows how planets are also formed outside of the initial planetesimal distribution. 

In this paper, we did not discuss the gas-disk model and the multi-simulation mode of GENGA as well as its built-in TTV tool used in \cite{Grimm+2018} for estimating the masses of the TRAPPIST-1 exoplanets. These topics will be improved and reported on in the future, as well as new tools or forces that will be be added to the present version of GENGA.

With the presented updates, it is now possible to run simulations with more than 30,000 fully interactive planetesimals in a feasible time. The added forces allow us to study the dynamics of small bodies in the solar system, or to study the formation and subsequent evolution of compact exoplanetary systems where tidal forces play an important role. It is also important that the field of high-performance computing and GPU computing evolves quickly and that simulation software must be constantly adapted and evolved to be able to use the most recent hardware efficiently. With this work, we aimed to provide a fast and user-friendly code for the study of terrestrial planet formation and evolution of the solar system and other exoplanetary systems.  

\section*{Acknowledgements}
Calculations were performed on UBELIX (http://www.id.unibe.ch/hpc), the HPC cluster at the University of Bern.

This work has been carried out within the framework of the NCCR PlanetS supported by the Swiss National Science Foundation.

CM acknowledges the support from the Swiss National Science Foundation under grant 200021\_204847 "PlanetsInTime".

We thank J{\o}rgen Nordmoen for running performance tests on the NVIDIA Tesla A100 and the AMD Instinct MI100 GPUs at the University of Oslo. 

\bibliographystyle{aasjournal}

\bibliography{genga3}


\section*{Appendices}
\begin{appendices}

\newpage
\section{Multi-level Close Encounter Algorithm}
\label{sect:algorithm}

The basic hybrid symplectic integrator can be written with the following scheme:

\begin{lstlisting}
Rcrit(dt)
Kick(dt/2), close-encounter pre-check
HC(dt/2)
FG(dt)
if(close-encounter candidates):
  close-encounter detection
  if(close-encounters):
    grouping
    Bulirsch-Stoer(dt)
HC(dt/2)
Kick(dt/2)
\end{lstlisting}

where \verb|Kick| corresponds to equation \ref{Hb} i.e. the interaction part of the Hamiltonian system, \verb|HC| to equation \ref{Hc} i.e. the democratic momentum summation, and \verb|FG| to equation \ref{Ha} i.e. the Kepler part of the Hamiltonian system. The parameter \verb|dt| is the time step. For more details on these routines we refer to \cite{GrimmStadel2014}.

To include the additional symplectic levels into the integrator, we replace the Bulirsch-Stoer call with a recursive function \verb|SEnc| (\textbf{S}ymplectic \textbf{Enc}ounters) and use the parameter \verb|SLevels| to indicate the number of symplectic levels.  We have

\begin{lstlisting}
Rcrit(dt)
Kick(dt/2), close-encounter pre-check
HC(dt/2)
FG(dt)
if(close-encounter candidates):
  close-encounter detection
  if(SLevels > 1):
    if(close-encounters):
      SEnc(dt, 1.0, 0)
  else:
    if(close-encounters):
      grouping
      Bulirsch-Stoer(dt)
HC(dt/2)
Kick(dt/2)
\end{lstlisting}

And the function \verb|SEnc| consists of:
\newpage
\begin{lstlisting}
SEnc(dt, ds, SLevel):
  groupS
  SLevel += 1
  ds *= substeps
  for(int s = 0; s < substeps; s++):
    RcritS()
    KickS(dt/ds / 2 )
    FGS(dt/ds)
    close-encounter detection
    if(close-encounters):
       if(SLevel < SLevels - 1):
         SEnc(dt, ds, SLevel)
       else:
        grouping
        Bulirsch-Stoer(dt/ds)
    KickS(dt/ds / 2 )
\end{lstlisting}

where \verb|substeps| is the number of sub-steps for each level, and $1/ds$ is the reduction factor of the time step. The \verb|groupS| function extracts all relevant close-encounter pairs from the upper symplectic level. It uses a scan function to perform a parallel stream compaction operation to generate the new close-encounter lists. The functions \verb|RcritS|, \verb|KickS|, and \verb|FGS| are similar to the original functions, but are applied only to the relevant bodies in each symplectic level.

\section{Additional new tools in GENGA}
\label{sect:AppB}
In this appendix, we list new tools in GENGA, which are not described in the main text. 

\subsection{Real-time visualization with openGL}
\begin{figure}
\includegraphics[width=1.0\linewidth]{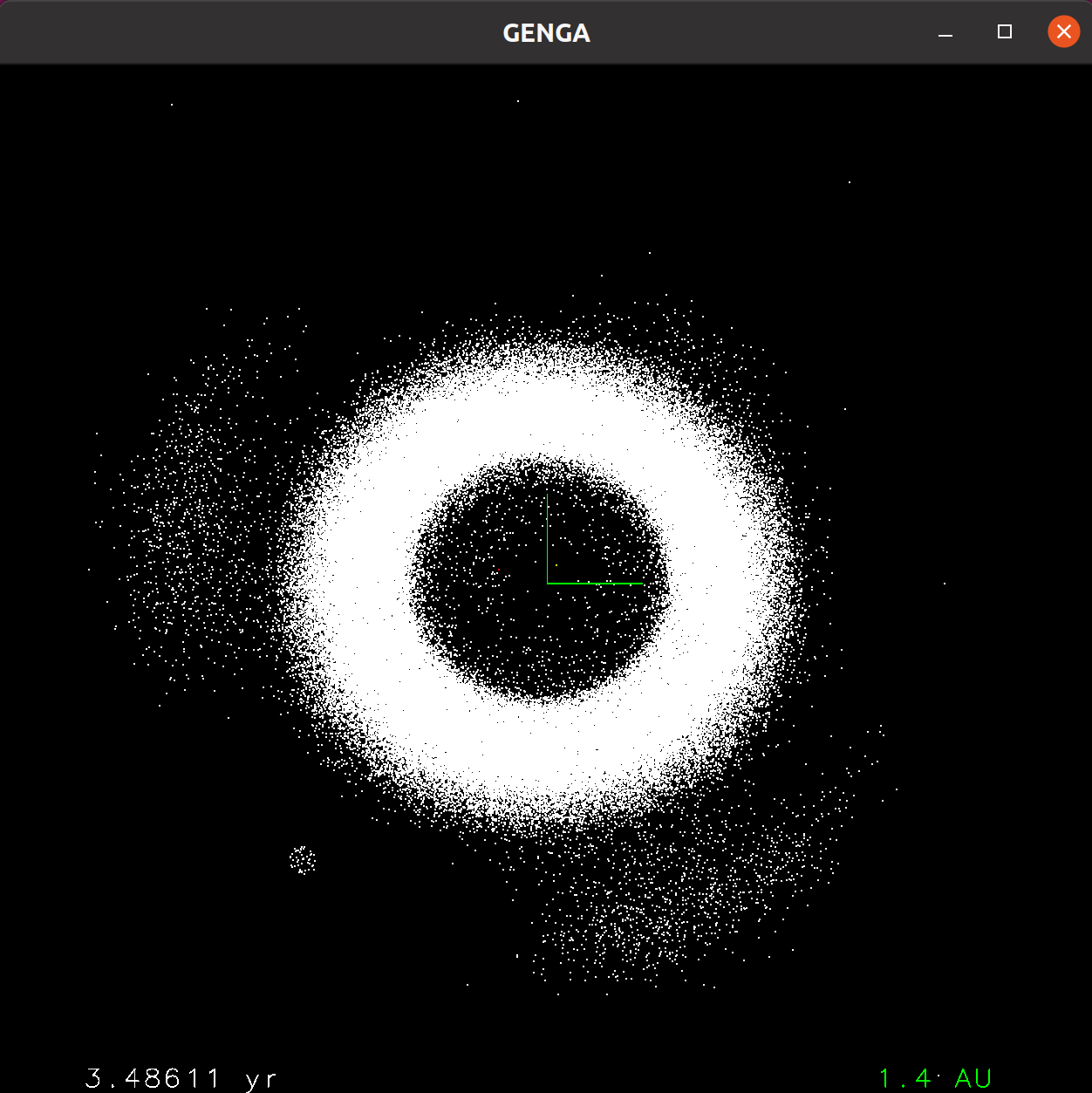}
\caption{Screenshot of the GENGA real-time visualization tool using openGL illustrating the inner part of the solar system, including Jupiter and the asteroid belt. The Trojans are nicely visible on the Lagrange points of Jupiter.}
\label{fig:gengaGL} 
\end{figure}

The Nvidia CUDA drivers offer the option to use openGL to visualize data on the GPU directly on the screen. Since no data transfer from the GPU to the CPU is needed, the visualization can be done very efficiently. We provide such an openGL interface for GENGA where a simulation can be projected in real time to the screen. Basic user interactions are possible, like zooming into the simulation or moving and rotating the viewing position. Having the possibility to experience the dynamics of a simulation in real time can help to build up a deeper intuition on the physics. Furthermore, short time dynamical effects such as satellite capture events can be observed in a simple and intuitive way. The only requirement to use the openGL interoperability is that the GPU must be connected to a monitor. A screenshot of the real-time visualization is shown in Figure \ref{fig:gengaGL}.

\subsection{Use calendar file for irregular output times}
When outputs are needed at irregular times, then GENGA offers an option to specify a calendar file, containing all desired output times. When the integration has reached an irregular time step, then the time step is reduced to the desired length and the output data is produced. After that, the reduced time step is reverted to get back to the last regular time step and the integration is continued as usual. In this way, no additional energy error is added.

\subsection{Exact reproducible results}
\label{sect:SerialGrouping}
Usually, the outcome of a parallel numerical calculation is not exactly reproducible. The reason is that calculations are done with a finite precision and that the outcome depends on the order of the operations. In general, we have $a + b + c \neq a + c + b$. Often the planetary $N$-body problem is chaotic, and therefore tiny differences in the calculation can lead to a different result on individual bodies. Even the number of formed planets can vary from simulation to simulation. A detailed study about this effect on GPUs is given in \cite{Hoffmann2017}.

However it is possible to force GENGA to exactly reproduce a given outcome. This is possible because on the GPU, summations of arrays are performed with parallel reduction sums. These parallel reduction sums have a well-defined order in which the terms are calculated, and the result is therefore always reproducible if the structure of the parallel reduction sum remains the same. The structure can, however, change by using different kernel parameters (see section \ref{sect:self-tuning}), or by using a different GPU type.

The only place that has not a fixed order in GENGA is the creation of the close-encounter pair lists. By introducing an additional sorting step on the close-encounter pairs lists, this order can be fixed and the result of a given initial condition is always the same. It is important to note that this does not mean that the results are 'true'; they still suffer from small round-off errors, just that this error is now always the same as before. Since the additional sorting step also introduces a performance penalty, it is not recommended to use this mode of GENGA for production runs. It is mostly useful to check if a GPU works correctly and it helps to eliminate memory leaks in a code. The reproducible option is only supposed to work on the same type of GPU, using different GPU types can still change the order of operations in the summation parts.

To enable the exact reproducible outcome mode in GENGA, the 'Serial Grouping' option must be used.

\subsection{The a-e and a-i grid option}
\begin{figure}
\includegraphics[width=1.0\linewidth]{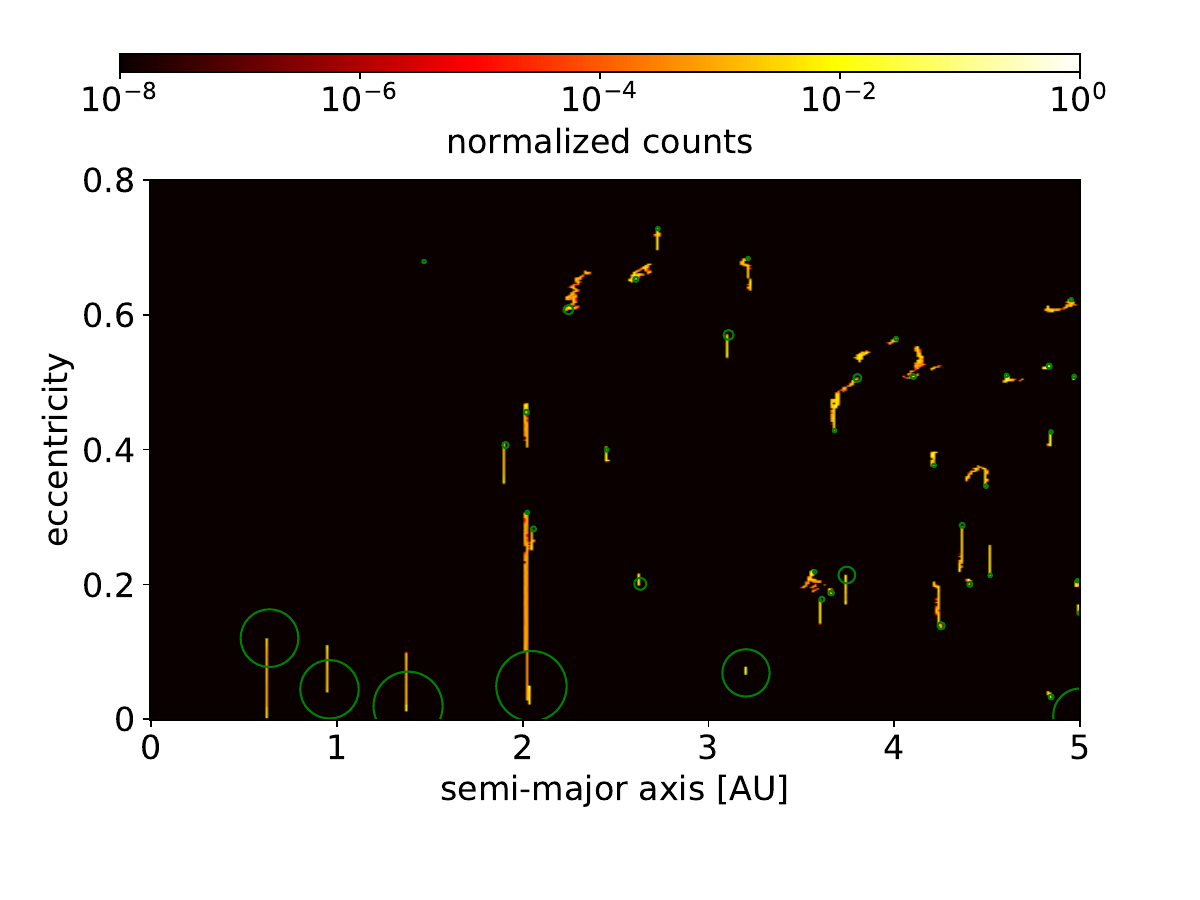}
\caption{The a-e grid as an example of planet formation. The green circles show a snapshot of the simulation where the semi-major axis and eccentricity represent a single moment in time. The size of the circles corresponds to the masses of the bodies. The color map on top of the green circles shows the a-e grid, where the positions of the bodies of all time steps of a given interval are included. The a-e grid is a two-dimensional histogram and allows to visualize short scale dynamics without using a lot of additional memory. 
}
\label{fig:aeCount} 
\end{figure}

The a-e and a-e grids are two-dimensional histograms. They count the number of particles in a given time span for each cell in a 'semi-major axis vs. eccentricity' or a 'semi-major axis vs. inclination' grid. The grids are calculated on the fly directly on the GPU, and they are particularly useful to visualize short time dynamics of a system without having to write a lot of output files. The grids are updated during the FG step. Since multiple bodies in parallel could contribute to the same a-e or a-i cell count, we use an atomicAdd\footnote{A CUDA atomic function performs its operation without interference from other threads and is therefore thread safe.} operation to make sure that every body is counted correctly.
This usage of the CUDA atomicAdd functions could potentially slow down the simulation when a lot of particles contribute to the same cell, but they also reduce the needed amount of memory. The a-e and a-i grids are stored in the GPU memory. Only at given intervals, the data is transferred back to the CPU and written to a file. In Figure \ref{fig:aeCount} is shown an example of the a-e grid. It can help to visualize the dynamics of a system, and makes short time movements of the particles visible.

\subsection{GPU Output Buffer}
When the state of a simulation needs to be written to a file, then first the data of all the bodies needs to be transferred back from the GPU to the CPU via the PCI-Express bus on the motherboard. When this data transfer is requested very frequently, then it can create a bottleneck of the entire run time. In order to increase the data transfer rate, we implement a GPU buffer that stores the output data temporarily on the GPU and then moves a bigger amount of data together, which can improve the performance. The size of this buffer can be set by the user in the GENGA parameter file. The GPU output buffer is especially useful in the multi-simulation mode when outputs are written at an interval of less than 10 time steps.

\end{appendices}

\end{document}